\documentclass[aps,onecolumn,showpacs]{revtex4}
\usepackage{dcolumn}
\usepackage{listings}
\usepackage{graphicx}
\usepackage{amsmath}
\usepackage{amsfonts}
\usepackage{amssymb}
\usepackage{psfrag}
\usepackage{wrapfig}
\usepackage{subfigure}
\usepackage{makeidx}
\usepackage{bm}
\usepackage{epsf}
\newtheorem{thm}{Theorem}
\newtheorem{lem}{Lemma}
\begin{document}

\title{A simple determinant representation for Rogue waves of Nonlinear Schr\"odinger equation}
\author{Liming Ling$^{1}$, Li-Chen Zhao$^{2}$}\email{zhaolichen3@163.com}
\address{$^1$Department of Mathematics, South China University of Technology, Guangzhou 510640, China
}
\address{$^2$Department of Physics, Northwest University, Xi'an
710069, China}

\date{\today}
\begin{abstract}

 We present a simple representation for arbitrary-order rogue wave solution and
 study on the trajectories of them explicitly. We find that the
 trajectories of two valleys on whole temporal-spatial distribution all look like
 ``X" shape for rogue waves. Additionally, we present some new types of high-order rogue wave structures, which could be helpful to realize the complex dynamics of rogue wave.

\end{abstract}
 \maketitle

\section{Introduction}
Rogue wave (RW) is localized both in space and time and depicts a
unique event which seems to appear from nowhere and disappear
without a trace\cite{V. Ruban,N.Akhmediev,C. Kharif,A. R.Osborne,E.
Pelinovsky}. Many
studies indicate that nonlinear theories can be used to explain the
dramatic phenomena \cite{Kibler,Chabchoub,Bailung}. Among nonlinear theories, the most fundamental
is based on the nonlinear Schr\"{o}dinger equation (NLS) \cite{R.
Osborne}
\begin{equation}\label{NLS}
i\frac{\partial u(x,t)}{\partial t}+\frac{\partial^2
u(x,t)}{\partial x^2}+2|u(x,t)|^2u(x,t)=0,
\end{equation}
which can be used to describe dynamics of localized waves in many physical systems, such as nonlinear fiber\cite{Kibler}, Bose-Einstein condensate\cite{matterrw}, plasma system\cite{Bailung}, and even water wave tank\cite{Chabchoub}. Particularly, the wave function describe the evolution of the electromagnetic field propagating in a nonlinear optics where the $z$ component takes the place of $t$
component \cite{NM}.  It can describe 1D ocean waves---focusing (defocusing) NLS
when dealing with deep (shallow) narrow banded water waves \cite{SS,KPS}.
The wave function can be the order parameter describing Bose-Einstein
condensates (NLS is called Gross-Pitaevskii equation in this field), for which defocusing NLS (repulsive interactions between atoms) is most of the
time used as the focusing case (attractive interactions between atoms)
in more than 1D leads to collapse events in the condensate \cite{PS}.

 The rational solution of the nonlinear equation has been used to describe the RW phenomena
\cite{Kibler,Chabchoub,Bailung}. It is now already more than 30 years since the first
breather solution of the NLS was found by Ma \cite{Ma}, which breathes temporally but is spatially
localized.
Akhmediev found a kind of
solutions \cite{Akhm,Akhm1} qualitatively different from Ma breathers, which were called by Akhmediev breather.
The Akhmediev breather breathe spatially but be localized in time.
Simply speaking, Akhmediev breathers are exact solutions
of the NLS that start from modulation instability of a plane
wave \cite{Akhm1} (also known as a Benjamin-Feir \cite{Ligh,Benj,Yuen} or
Bespalov-Talanov instability) and return to a plane wave
at the end of the evolution. Peregrine gave a solution localized in
both space and time in 1983 \cite{Pere}, which can be seen as the limit of Ma breather and Akhmediev breather. Recently, the Peregrine rational solution (fundamental RW), the second-order RW, and solutions up to order 5 have been observed in
a water wave tank \cite{Chabchoub,Chab2,Chab3,OPCK}. Also, the rogue waves described by Peregrine rational solution have
been generated in optics \cite{Kibler,Erk} and magnetoplasma \cite{Shu,Sabry}.
Thus, the validity of the simplest RW solutions has
been confirmed experimentally. This also means that the simple presentation and the quality for RW solutions
is crucial for the further research in this area.

Darboux transformation, originating from the work
of Darboux in 1882 on the Sturm-Liouville equation, is a
powerful method for constructing solutions for integrable
systems. The theory is presented in several monographs and
review papers (see \cite{Mat,Cies,Dok}). Various approaches
have been proposed to find a Darboux transformation for a
given equation, for instance, the operator factorization method
\cite{Adler}, the gauge transformation method \cite{Dok,Neu,Li}, and the loop
group transformation \cite{Terng}. The classical Darboux transformation can be used to derive fundamental RW solution. However, it can not
be applied to derive high-order RW solutions since the classical Darboux transformation can not be iterated at the same spectral parameter.
To overcome this difficulty, Guo, Liu and the first author of this paper used the limit technique to generalize the classical Darboux transformation \cite{Ling,GLL1}, which can be used to yield high-order solution. Based on this simple idea, there are a series of research papers about
high order RW solution for other integrable system \cite{He,He1}. It should be pointed that based on direct recursive Darboux transformation, numeric method and limit technique can be used to construct high order rogue wave solution \cite{KAA,KAA1,KAA2,KAA3}. Comparing with the previous method, we can give a simple representation for general high order RW solutions. Besides Darboux related method, there are also some other methods to derive general high-order RW solutions \cite{Yang,Matveev,Gaillard,Kalla}.

 In this paper, we present a simple representation for general NLS
 RW solution, and investigate the dynamics and kinetics of  RW
 explicitly. We find the whole trajectories for high-order RWs are similar to the one of
 the first-order RW, whose trajectory looks like an ``X". They are different around the location where the RW happens. Besides, we discuss the classification of them
 by parameters $s_i$. We present some new structures for general fourth-order rogue wave, such as ``double column" structure and ``claw-line" structure.

\section{A simple representation for general RW solution}
In this section,
we derive a generalized expression for arbitrary Nth-order RW
solution of Eq. \eqref{NLS}. The Lax pair for
NLS equation \eqref{NLS} is \cite{AKNS}:
\begin{equation}\label{Lax}
    \begin{split}
      \Psi_x &=(-i\lambda \sigma_3+iQ)\Psi, \\
      \Psi_t &=(2i\lambda^2\sigma_3-2i\lambda Q-i\sigma_3
      Q^2+\sigma_3Q_x)\Psi,
    \end{split}
\end{equation}
where
\begin{equation*}
    \sigma_3=\begin{pmatrix}
               1 & 0 \\
               0 & -1 \\
             \end{pmatrix},\quad Q=\begin{pmatrix}
                                     0 & \bar{u} \\
                                     u & 0 \\
                                   \end{pmatrix}.
\end{equation*}

 Previously we give some lemmas about generalized Darboux transformation.
\begin{lem}[\cite{Ling,GLL1},Guo, Ling and Liu]
Suppose we have $N$ different solutions $\Phi_i$ for system \eqref{Lax} with $\lambda=\lambda_i$,
then the N-fold Darboux transformation
\begin{equation*}
    \begin{split}
      T_N &=I+\sum_{i=1}^N\frac{T_i}{\lambda-\bar{\lambda}_i}  \\
          &=I-XM^{-1}(\lambda-D)^{-1}X^{\dag},
    \end{split}
\end{equation*}
where
\begin{equation*}
    \begin{split}
      M &= \left(\frac{\Phi_i^{\dag}\Phi_j}{\lambda_j-\bar{\lambda}_i}\right)_{1\leq i,j\leq N}, \\
      X &=[\Phi_1,\Phi_2,\cdots,\Phi_N], \\
      D &=\mathrm{diag}(\bar{\lambda}_1,\bar{\lambda}_2,\cdots,\bar{\lambda}_N),
    \end{split}
\end{equation*}
converts system \eqref{Lax} into new system
\begin{equation*}
    \begin{split}
      \Psi[N]_x &=(-i\lambda \sigma_3+iQ[N])\Psi[N], \\
      \Psi[N]_t &=(2i\lambda^2\sigma_3-2i\lambda Q[N]-i\sigma_3
      Q[N]^2+\sigma_3Q[N]_x)\Psi[N],
    \end{split}
\end{equation*}
where
\begin{equation*}
 Q[N]=\begin{pmatrix}
                                     0 & \overline{u[N]} \\
                                     u[N] & 0 \\
                                   \end{pmatrix},\quad u[N]=u-2\sum_{i=1}^NT_i[2,1],
\end{equation*}
\begin{equation*}
\begin{split}
&\sum_{i=1}^NT_i[2,1]=\frac{\det(M_1)}{\det(M)},\quad M_1=\begin{pmatrix}
          M & (M[1])^{\dag} \\
          M[2] & 0 \\
        \end{pmatrix},  \\
&M[i]=[\Phi_1[i],\Phi_2[i],\cdots,\Phi_N[i]],
\end{split}
\end{equation*}
$\Phi_m[i]$ is the $i$-th component of vector $\Phi_m$, $m=1,2,\cdots,N$, $i=1,2$.
\end{lem}
As we known from reference \cite{Ling}, the generalized Darboux transformation is noting but special limit for N-fold Darboux transformation. Based on above lemmas, we can obtain the following theorem to general high order rogue wave solution.
\begin{thm}
The general RW solution formula of NLS is
\begin{equation}\label{NLS-formula}
    u(x,t) =\left(1-2\frac{\det(A_1)}{\det(A)}\right)\exp(2it),
\end{equation}
and its density expression can be derived as
\begin{equation}\label{q2}
    |u|^2=1+\left[\ln{\det(A)}\right]_{xx} .
\end{equation}
The expressions for $A_1$, and $A$ are
\begin{eqnarray}
    A_1&=&\begin{pmatrix}
          A &A[2]  \\
          A[1] & 0 \\
        \end{pmatrix}
    ,\nonumber\\
A&=&(A_{l,j})_{1\leq l,j\leq N},
\end{eqnarray}
where
\begin{equation*}
A[1]=\left[\phi_0,\phi_1,\cdots \phi_{N-1}\right],\quad
    A[2]=\left[\bar{\psi_0},\bar{\psi_1},\cdots
    \bar{\psi}_{N-1}\right]^{T},
\end{equation*}
The variable function $\psi_j$, $\phi_j$ and $A_{i,j}$ are the
related Taylor expansion coefficients of the following functions
\begin{eqnarray}
\psi&=&i
(C_1X-C_2X^{-1})\equiv\sum_{i=0}^{+\infty}\psi_if^{2i},\nonumber\\
\phi&=&C_2X-C_1X^{-1}\equiv\sum_{i=0}^{+\infty}\phi_if^{2i},\nonumber\\
A(f,\bar{f})&=&\frac{
    i(\psi\bar{\psi}+\phi\bar{\phi})}{2+f^2+\bar{f}^{2}}=\sum_{l,j=1}^{+\infty,+\infty}A_{l,j}f^{2(l-1)}\bar{f}^{2(j-1)},\\
 \psi_i&=&\frac{1}{(2i)!}\frac{\partial^{2i} \psi}{\partial f^{2i}}|_{f=0},\quad \phi_i=\frac{1}{(2i)!}\frac{\partial^{2i} \phi}{\partial f^{2i}}|_{f=0},\nonumber\\
 A_{l,j}&=&\frac{1}{(2(l-1))!(2(j-1))!}\frac{\partial^{2(l+j-2)} A(f,\bar{f})}{\partial f^{2(l-1)}\partial \bar{f}^{2(j-1)}}|_{f=0}\nonumber
\end{eqnarray}
where $C_1=\frac{\sqrt{1+f^2-h}}{h}$,
$C_2=\frac{\sqrt{1+f^2+h}}{h}$, $X=e^{h[x+2it+2i t f^2+S(f)]}$,
$h=f\sqrt{2+f^2}$, and $S(f)=\sum_{i=1}^{N-1}s_if^{2i}$( $s_i$ is a
complex constant). The symbol ~$\bar{}$ represents the complex
conjugation. \end{thm} \textbf{Proof:} We prove this theorem based
on generalized Darboux transformation. We neglect the proof for the
generalized Darboux transformation, since the details are given in
references \cite{Ling,GLL1,Mat,Terng,ZM}.

From the seed solution for NLS equation \eqref{NLS}
$u[0]=\exp[2it]$,  we have the fundamental solution for Lax pair
\eqref{Lax}
\begin{eqnarray}
    \Psi_0&=&\exp[-it\sigma_3]
    \begin{pmatrix}
    1 & \sqrt{1+\lambda^2}-\lambda \\
    \lambda-\sqrt{1+\lambda^2} & 1 \\
    \end{pmatrix} \exp[-i(1+\lambda^2)^{1/2}(x-2\lambda t)\sigma_3].
\end{eqnarray}

Together with the generalized Darboux transformation $T$ \cite{Ling,GLL1}, we can
obtain the fundamental solution matrix $\Psi=T\Psi_0$ for general
rogue wave solution $u$. Furthermore, the solution matrix can be expanded in following form
\begin{equation*}
    \Psi=E\exp[-i\lambda \sigma_3(x-2\lambda t)],\quad
    \text{as } \lambda\rightarrow \infty^{+},
\end{equation*}
here $E=I+E_1\lambda^{-1}+E_2\lambda^{-2}+O(\lambda^{-3}).$ Substituting into \eqref{Lax},  we
have
\begin{equation*}
    \begin{split}
      E_x &=-i\lambda[\sigma_3,E]+iQE, \\
      E_t  &=2i\lambda^2[\sigma_3,E]+(-2i\lambda Q-{\rm
      i}\sigma_3Q^2+\sigma_3Q_x)E.
    \end{split}
\end{equation*}
Comparing the coefficient of above two equations, then we have
\begin{equation*}
    \begin{split}
      &Q =[\sigma_3,E_1],  \\
      &E_{1,x}=-i[\sigma_3,E_2]+iQE_1,\\
      &2i[\sigma_3,E_2]-i\sigma_3Q^2+\sigma_3Q_x-2{\rm
      i}QE_1=0.
    \end{split}
\end{equation*}
It follows that
\begin{equation*}
    \sigma_3Q^2=2iE_{1,x}^{diag}.
\end{equation*}

To realize the above expansion form, the Taylor expansion for $ \sqrt{1+\lambda^2}$ is essential:
\begin{equation*}
    \sqrt{1+\lambda^2}=\lambda+\frac{1}{2}\lambda^{-1}-\frac{1}{8}\lambda^{-3}+O(\lambda^{-5}).
\end{equation*}
Together with the Darboux transformation\cite{Ling,GLL1}
\begin{equation*}
    T=I+\sum_{i=1}^N\frac{T_i}{\lambda-\bar{\lambda}_i},
\end{equation*}
then the exact expression of $E_1$ is
\begin{equation*}
    E_1=-\frac{ix}{2}\sigma_3+\frac{1}{2}\begin{pmatrix}
                                           0 & e^{-2it} \\
                                           -e^{2it} & 0 \\
                                         \end{pmatrix}+\sum_{i=1}^NT_i.
\end{equation*}
It follows that the exact formula for $u$ and $|u|^2$ can be obtained as:
\begin{equation*}
\begin{split}
  u &=\exp(2it)-2\sum_{i=1}^N T_i[2,1], \\
  |u|^2&=1+2i\left(\sum_{i=1}^N T_i[1,1]\right)_x\\
  &=1-2i\left(\sum_{i=1}^N T_i[2,2]\right)_x,\\
  &=1+i\left(\sum_{i=1}^N T_i[1,1]-T_i[2,2]\right)_x,
\end{split}
\end{equation*}
where $T_i[m,n]$ represents the $(m,n)$
element of matrix $T_i$. On the other hand, we have
\begin{equation*}
    \left(\frac{\Phi_i^{\dag}\Phi_j}{\lambda_j-\bar{\lambda}_i}\right)_x=-i \Phi_i^{\dag}\sigma_3\Phi_j.
\end{equation*}
Together with above lemma,
the exact rogue wave
solution can be obtained by the limit technique \cite{Ling,GLL1}.
$\square$

The formula $A_{l,j}$ can be rewritten as
\begin{equation}\label{Aij}
    A_{l,j}=\frac{i}{2}\sum_{m=0, \alpha\leq j-1\leq l+j+\alpha-m-2,}^{l+j-2}\sum_{\alpha=0,0\leq \alpha\leq m}^{j-1}(-\frac{1}{2})^{l+j-m-2}C^{j-1-\alpha}_{l+j-2-m}(
    \psi_{m-\alpha}\bar{\psi}_{\alpha}+\phi_{m-\alpha}\bar{\phi}_{\alpha}),
\end{equation}
where $C_m^n=\frac{m!}{n!(m-n)!}.$
Indeed, we can prove the above equality by the following Taylor expansion
\begin{equation*}
    \begin{split}
    &\frac{i(\psi\bar{\psi}+\phi\bar{\phi})}{2+f^2+\bar{f}^{2}}\\
        =&\frac{i}{2}\left[\sum_{i=0}^{\infty}\left(\sum_{j=0}^i(\psi_{i-j}\bar{\psi}_{j}+\phi_{i-j}\bar{\phi}_{j})f^{2(i-j)}\bar{f}^{2j}\right)\right]\cdot \left[\sum_{j=0}^{+\infty}(-\frac{1}{2})^j(f^2+\bar{f}^2)^j\right] \\
       =& \frac{i}{2}\sum_{i=0}^{+\infty}\sum_{m=0}^i(-\frac{1}{2})^{i-m}(f^2+\bar{f}^2)^{i-m}
       \left(\sum_{j=0}^m(\psi_{m-j}\bar{\psi}_{j}+\phi_{m-j}\bar{\phi}_{j})f^{2(m-j)}\bar{f}^{2j}\right)\\
       =&\frac{i}{2}\sum_{i=0}^{+\infty}\sum_{m=0}^i\left[\left(\sum_{j=0}^m(\psi_{m-j}\bar{\psi}_{j}+\phi_{m-j}\bar{\phi}_{j})f^{2(m-j)}\bar{f}^{2j}\right)
       (-\frac{1}{2})^{i-m}\cdot \sum_{\alpha=0}^{i-m}C_{i-m}^{\alpha}f^{2(i-m-\alpha)}\bar{f}^{2\alpha}\right]\\
       =&\frac{i}{2}\sum_{i=0}^{+\infty}\sum_{m=0}^i\left[(-\frac{1}{2})^{i-m}\sum_{l=0,j\leq l\leq i-m+j,}^i\sum_{j=0,0\leq j\leq m}^lC^{l-j}_{i-m}(
    \psi_{m-j}\bar{\psi}_{j}+\phi_{m-j}\bar{\phi}_{j})f^{2(i-l)}\bar{f}^{2l}\right]\\
    =&\frac{i}{2}\sum_{i=0}^{+\infty}\sum_{l=0}^i\left[\sum_{m=0,j\leq l\leq i-m+j,}^i\sum_{j=0,0\leq j\leq m}^l(-\frac{1}{2})^{i-m}C^{l-j}_{i-m}(
    \psi_{m-j}\bar{\psi}_{j}+\phi_{m-j}\bar{\phi}_{j})\right]f^{2(i-l)}\bar{f}^{2l}.
    \end{split}
\end{equation*}
A transformation $i-l\rightarrow l-1$, $l\rightarrow j-1$ could
obtain the formula \eqref{Aij}.

\section{The application of theorem 1 and the trajectories of Rogue waves}
In order to illustrate how to use the above theorem, we give the following examples. Taking parameters $s_1=a+ib$, $s_2=c+id$, $s_j=0$, $j\geq3$,  we have the following explicit expression
\begin{equation*}
    \begin{split}
     \psi_0=&2 i x-4 t-i,\\
     \psi_1=&\left[\frac{2}{3}i x^3-(4 t+i) x^2+\left(-8 i t^2+4 t+\frac{1}{2} i\right) x+\frac{16}{3} t^3+4 i t^2-5 t+\frac{1}{4} i-2 b+2 i a\right],\\
     \psi_2=&\left[\frac{1}{15} i x^5-\left(\frac{2}{3} t+\frac{1}{6} i\right) x^4+\left(-\frac{8}{3}i t^2+\frac{4}{3} t +\frac{1}{2} i\right) x^3+\left(+\frac{16}{3} t^3+4 it^2-7 t-\frac{1}{4} i+2 i a-2 b\right) x^2\right.\\
     &+\left(\frac{16}{3} i t^4-\frac{16}{3} t^3-22 i t^2+(5-8 i b-8 a) t-\frac{1}{16} i+2 b-2 i a\right) x-\frac{32}{15} t^5-\frac{8}{3} i t^4+20 t^3\\
     &\left.+(-8 i a+8 b+9 i) t^2+\left(4 a+4 i b-\frac{7}{8}\right) t-2 d+2 i c-\frac{1}{2}b+\frac{1}{2} i a-\frac{3}{32} i\right],
    \end{split}
\end{equation*}
and
\begin{equation*}
    \begin{split}
      \phi_0=&2 x+4 i t+1,\\
      \phi_1=&\left[\frac{2}{3} x^3+(4 i t+1) x^2+\left(-8 t^2+4 i t+\frac{1}{2}\right) x-\frac{16}{3} i t^3-4 t^2+5it-\frac{1}{4}+2ib+2 a\right],\\
      \phi_2=&\left[\frac{1}{15} x^5+\left(\frac{2}{3} i t+\frac{1}{6}\right) x^4+\left(-\frac{8}{3} t^2+\frac{4}{3} i t+\frac{1}{2}\right) x^3+\left(-\frac{16}{3} i t^3-4 t^2+7 i t+2 a+2 i b+\frac{1}{4}\right) x^2\right.\\
      &+\left(\frac{16}{3} t^4-\frac{16}{3} i t^3-22 t^2+(5 i-8 b+8 i a) t+2 a+2 i b-\frac{1}{16}\right) x+\frac{32}{15} i t^5+\frac{8}{3} t^4-20 i t^3\\
      &\left.+(-9-8 i b-8 a) t^2+\left(4 ia+\frac{7}{8} i-4 b\right)t+2 c+\frac{1}{2} a+2id+\frac{1}{2} i b+\frac{3}{32}\right].
    \end{split}
\end{equation*}
Together with the formula \eqref{Aij}, we have the following
explicit expression $A_{l,j}$:
\begin{equation*}
\begin{split}
   A_{11}=&\frac{{\rm i}}{2}(\psi_0\bar{\psi_0}+\phi_0\bar{\phi_0})= i \left( 1+4\,{x}^{2}+16\,{t}^{2} \right)   \\
   A_{12}=&\frac{{\rm i}}{2}\left[-\frac{1}{2}(\psi_0\bar{\psi_0}+\phi_0\bar{\phi_0})+(\psi_1\bar{\psi_0}+\phi_1\bar{\phi_0})\right]
\\
  =&\frac{1}{2}\left( \left( 16\,t+8\,ix \right) a+ \left( 16\,it-8\,x \right) b+\frac{8}{3}\,i{x}^
{4}-{\frac {32}{3}}\,{x}^{3}t+ \left( -{\frac
{128}{3}}\,{t}^{3}-24\,t
 \right) x-\frac{3}{2}\,i+16\,i{t}^{2}-{\frac {128}{3}}\,i{t}^{4}
\right)   \\
  A_{21}=&\frac{{\rm i}}{2}\left[-\frac{1}{2}(\psi_0\bar{\psi_0}+\phi_0\bar{\phi_0})+(\psi_0\bar{\psi_1}+\phi_0\bar{\phi_1})\right] \\
  =&\frac{1}{2}\left(\left( -16\,t+8\,ix \right) a+ \left( 8\,x+16\,it \right) b+\frac{8}{3}\,i{x}
^{4}+{\frac {32}{3}}\,{x}^{3}t+ \left( {\frac
{128}{3}}\,{t}^{3}+24\,t
 \right) x-\frac{3}{2}\,i+16\,i{t}^{2}-{\frac {128}{3}}\,i{t}^{4}
\right)  \\
  A_{22}=&\frac{{\rm i}}{2}\left[\frac{1}{4}C_2^1(\psi_0\bar{\psi_0}+\phi_0\bar{\phi_0})-\frac{1}{2}\left[C_1^1(\psi_0\bar{\psi_1}+\phi_0\bar{\phi_1})
  +C_1^0(\psi_1\bar{\psi_0}+\phi_1\bar{\phi_0})\right]+(\psi_1\bar{\psi_1}+\phi_1\bar{\phi_1})\right]\\
  =&\frac{1}{4}\left[16\,i{a}^{2}+ \left( {\frac {32}{3}}\,i{x}^{3}+ \left( -128\,i{t}^{2}-
8\,i \right) x \right) a+16\,i{b}^{2}+ \left( -{\frac
{256}{3}}\,i{t}^ {3}+64\,i{x}^{2}t+48\,it \right) b
\right.\\&\left.+{\frac {16}{9}}\,i{x}^{6}+ \left(
\frac{4}{3}\,i+{\frac {64}{3}}\,i{t}^{2} \right) {x}^{4}+ \left(
{\frac {256}{3 }}\,i{t}^{4}+160\,i{t}^{2}-i \right) {x}^{2}+{\frac
{13}{ 4}}\,i+76\,i{t}^{2}-64\,i{t}^{4}+{\frac
{1024}{9}}\,i{t}^{6}\right]
\end{split}
\end{equation*}
 Furthermore, by formula \eqref{q2} we can
obtain that
\begin{equation*}
    |u[1]|^2=1+[\ln(A_{11})]_{xx}=1+[\ln(1+4x^2+16t^2)]_{xx},
\end{equation*}
\begin{equation*}
    \begin{split}
   |u[2]|^2=1+\left[\ln \begin{vmatrix}
                          A_{11} & A_{12} \\
                          A_{21} & A_{22} \\
                        \end{vmatrix}
   \right]_{xx}=1+\left[\ln \left(D_2\right)\right]_{xx},
    \end{split}
\end{equation*}
where
\begin{equation}\label{D2}
\begin{split}
  D_2&=144 a^2+
    (144 x-192 x^3+2304 x t^2) a+144 b^2\\
    &+(-1152 x^2 t+864 t+1536
    t^3)b
   +9+108 x^2+1584 t^2\\
   &+48 x^4+6912 t^4+768 x^4 t^2+64 x^6-1152 x^2 t^2\\
   &+4096 t^6+3072 x^2
   t^4.
\end{split}
\end{equation}
Similar as above, we can obtain that
\begin{equation*}
    \begin{split}
   |u[3]|^2=1+\left[\ln\left(\sum_{j=0}^{12}F_{j}x^{j}\right)\right]_{xx},
    \end{split}
\end{equation*}
the explicit expression of $F_j$ is giving on the Appendix.

We illustrate that we give a simple formula of general rogue wave
solution for NLS equation \eqref{NLS}. Reference \cite{N.A} was the
original work giving the first five orders of rogue waves, a
modified Darboux transformation was used. Many different methods
have been performed to derive Nth-order RW solution, such as the
Hirota bilinear method \cite{Yang} , reduction method
\cite{Matveev}, algebraic geometry solution reduction method
\cite{Gaillard,Kalla}, and the generalized Darboux transformation
method \cite{Ling,He}. However, the formulas presented in these
papers are all the ratio of two $2N\times 2N$ determinants. In this
paper, we develop the generalized Darboux transformation method
\cite{Ling} to present a much simpler representation for Nth-order
RWs with ratio of $(N+1)\times(N+1)$ order determinant and $N\times
N$ order determinant. The $N$ order and $2N$ order determinants have
$N!$ terms and $(2N)!$ terms respectively. As $N$ big enough, the
terms of $2N$ order determinant are much more that $N$ order. Such
as $N=10$, we have $10!=3628800$, $11!=39916800$,
$20!=2432902008176640000$. Thus our formula  simplifies the
calculation tremendously. In addition, by the formula \eqref{q2}, we
can see the conservation law
$$\int_{-\infty}^{+\infty}(|u|^2-1) dx=\frac{[\det(A)]_x}{\det(A)}|_{x=-\infty}^{x=+\infty}=0,$$
since $\det(A)$ is a polynomial of $x$.

The fundamental RW solution has been given for a long time
\cite{Pere}, and its explicit formation can be given directly
from the generalized expression \eqref{NLS-formula} with $N=1$ as
\begin{equation}\label{u1}
    u[1]=\left(1-\frac{2
    }{A_{11}}\begin{vmatrix}
                          A_{11} & \bar{\psi_0} \\
                          \phi_0 & 0 \\
                        \end{vmatrix}\right)=\left(-1+\frac{4(1+4it)}{1+4x^2+16t^2}\right)e^{2it}
\end{equation}

\begin{figure}[htb]
\centering
\subfigure[]{\includegraphics[height=45mm,width=50mm]{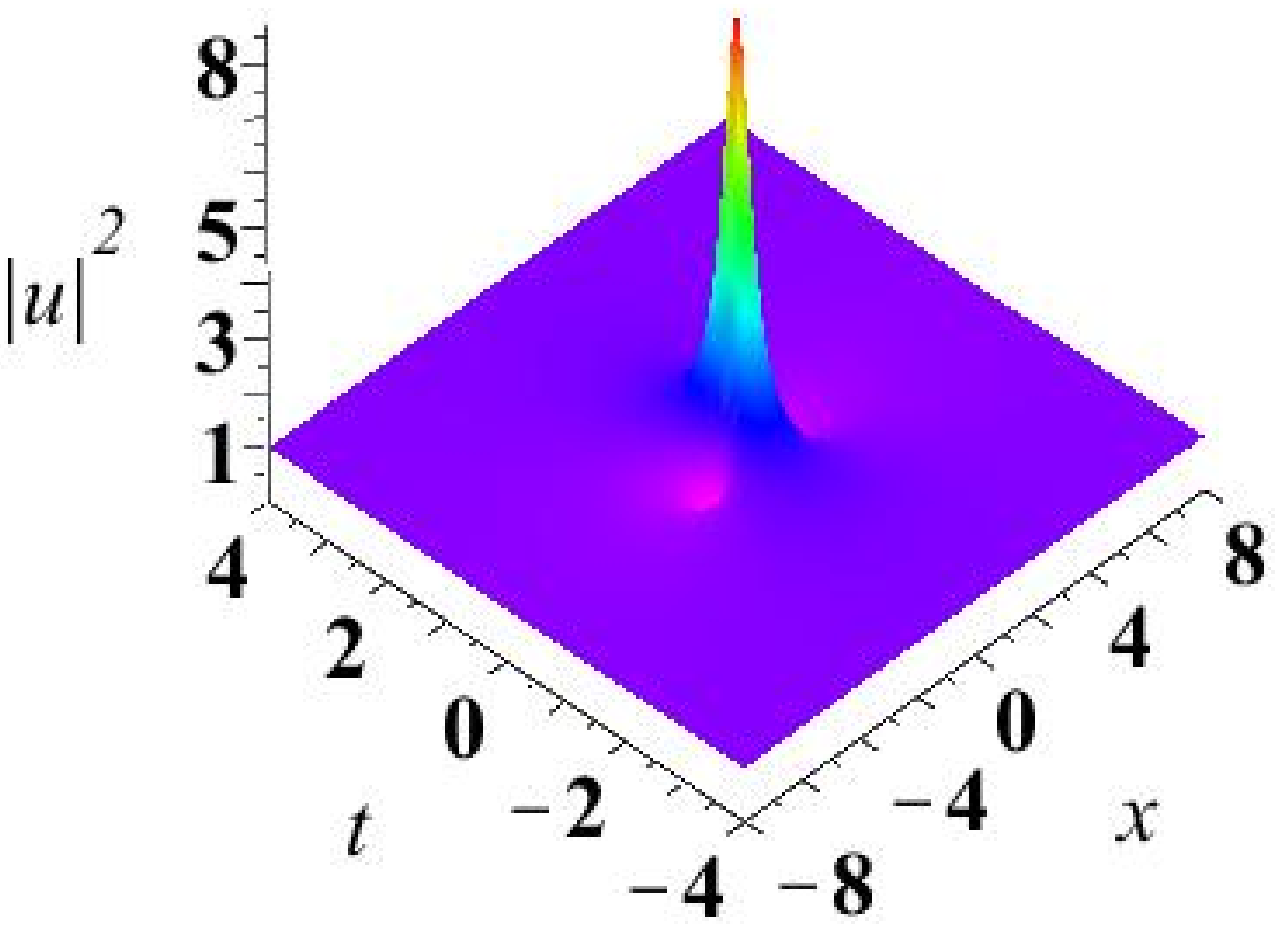}}
\hfil
\subfigure[]{\includegraphics[height=35mm,width=35mm]{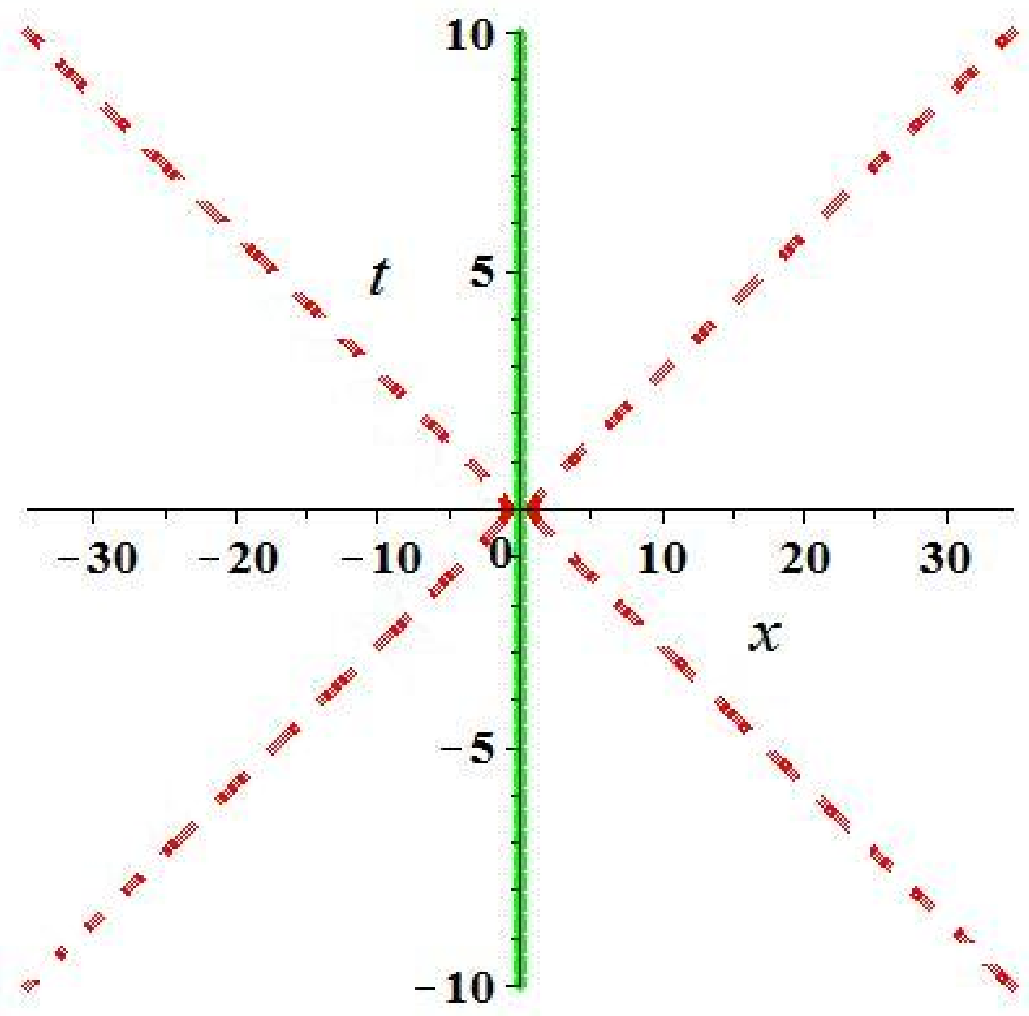}}
\caption{(color online) (a) The
density evolution of the first-order RW.  (b) The trace of the first-order RW. The green solid line
corresponds to the trajectory of the RW's hump and the red dashed
lines correspond to the trajectories of the RW's valleys. This holds
for all pictures in the paper.}\label{fig1}
\end{figure}
 The solution corresponds to the well-known ``eyes" shaped RW. Near
 $t=0$, the wave has highest hump and there are two valleys around
 the hump (see Fig. \ref{fig1}(a)). Long before or after $t=0$, the
 peak values of the hump and valley are close to the background.
 But the wave keep the structure
 before or after the moment $t=0$.  Therefore, we can define the
 trajectory by the motion of its hump and the valleys, which can be
 described by the motion of the hump and valleys' center locations
  \cite{Zhao}. The
motion of its hump's center is calculated as
\begin{equation}\label{hump}
X_h=0,
\end{equation}
and the motions of the two valleys' center are
\begin{equation}\label{vallley}
X_v=\pm\sqrt{3}(t^2+\frac{3}{48})^{1/2}.
\end{equation}
Then, we can plot the RW's trajectory in Fig. \ref{fig1} (b). The trajectories of the two valleys
look like an ``X" shape,  as shown by the red dashed lines in Fig. \ref{fig1}(b).  The
trajectory of RW's hump is a straight line which traverse the center of the  ``X" . Moreover, the straight line is one of the symmetric axis of  the  ``X".

 Furthermore, we can define the
width of RW as the distance between the two valleys' centers, which
corresponds to the distance between the two red lines in Fig. \ref{fig1}(b).
Its evolution is
\begin{equation}
W=2 \sqrt{3}(t^2+\frac{3}{48})^{1/2}.
\end{equation}
Obviously, the width is compressed before the moment $t=0$ when the
highest peak emerge and is broadened after the moment.

\begin{figure}[htb]
\centering
\subfigure[]{\includegraphics[height=30mm,width=40mm]{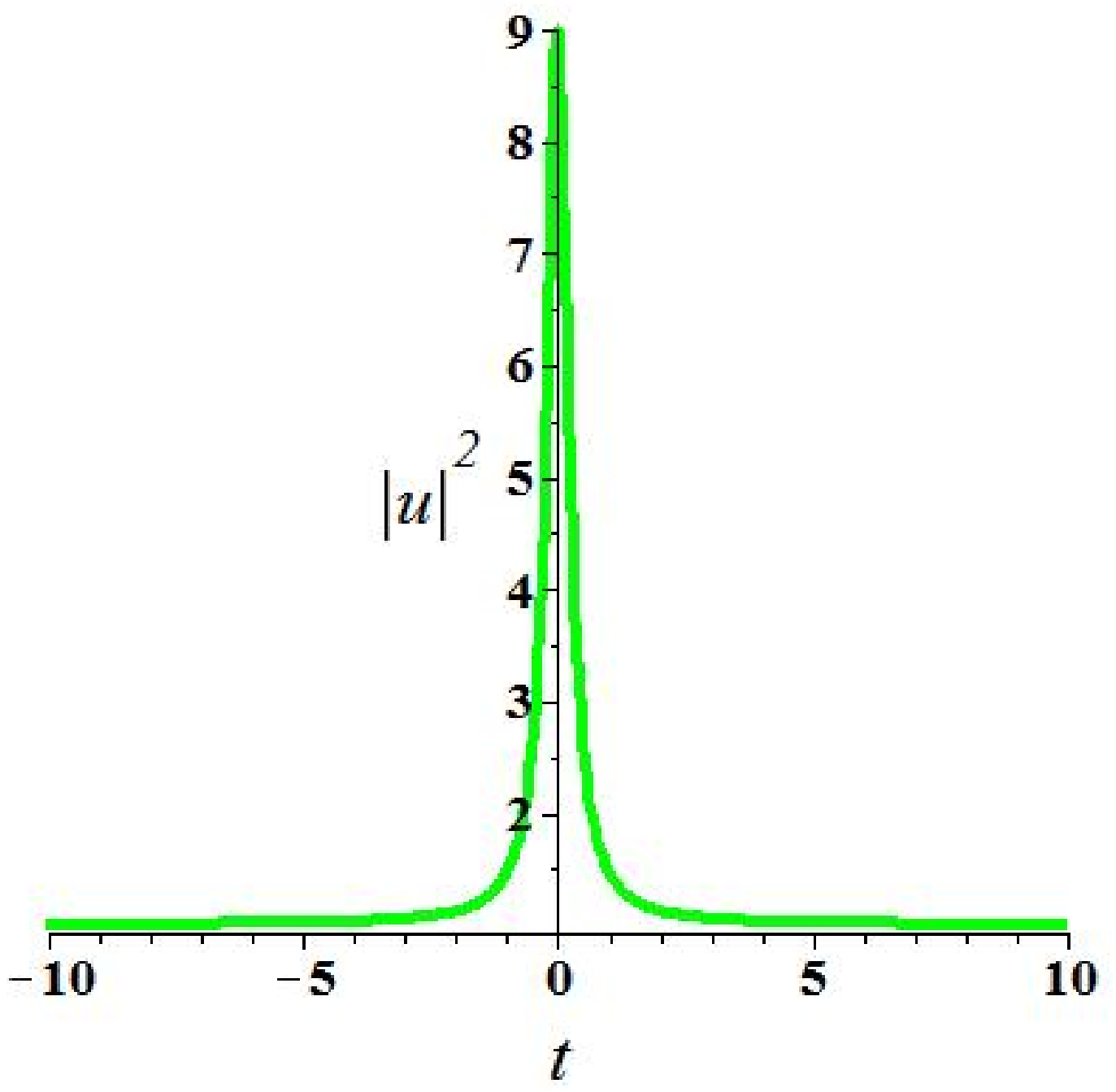}}
\hfil
\subfigure[]{\includegraphics[height=30mm,width=40mm]{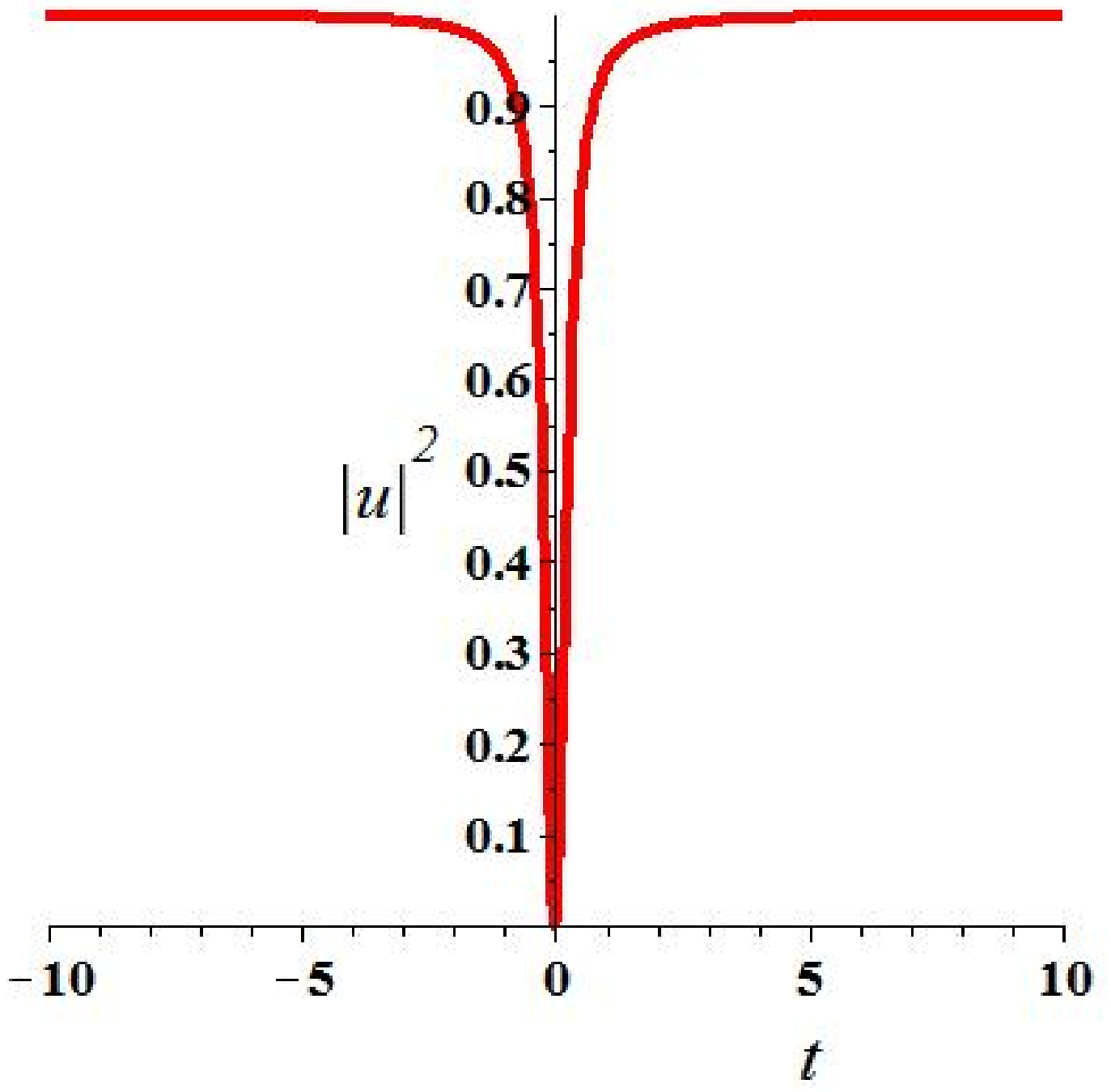}}
\caption{(color online) (a) The value evolution of RW's hump. (b)
The value evolution of RW's valleys. }\label{fig2}
\end{figure}

Inserting the \eqref{hump} and \eqref{vallley} into the density expression of the \eqref{u1} , one can obtain explicit expressions for the evolution of RW's hump and valleys, shown in Fig. \ref{fig2}. It
is seen that the highest value of the RW is nine times the value of
background\cite{N.A}. Compare Fig. \ref{fig1}(b) and Fig. \ref{fig2}, we know that RW has the
highest and steepest structure when the width is the smallest.
Namely, the width is broaden or compressed with time, the peak
 decrease or increase correspondingly from the modulation instability effect.

Higher-order RW solution has been presented in \cite{Ling,Yang,N.A}.
It is known that the second order rogue wave possesses different
dynamics. Choosing the parameter $s_1=a+ib$, we can readily obtain
the general second order rogue wave solution by the formula
\eqref{NLS-formula}
\begin{equation*}
    u[2]=\left(1-\frac{2}{\begin{vmatrix}
                            A_{11} & A_{12} \\
                            A_{21} & A_{22} \\
                          \end{vmatrix}
    }\begin{vmatrix}
                            A_{11} & A_{12}&\bar{\psi_0} \\
                            A_{21} & A_{22}&\bar{\psi_1} \\
                            \phi_0 & \phi_1&0 \\
                          \end{vmatrix}\right)=\left(1+\frac{D_1}{D_2}\right)e^{2it},
\end{equation*}
\begin{eqnarray}
D_1&=&(-576 x-2304 i x t) a+(144 i-2304 i t^2-1152
t \nonumber\\
&&+576 i x^2) b-768 i x^4 t-192 x^4
    -1536 i t^3
    -288 x^2\nonumber\\
&&-3456 t^2+36-4608 x^2 t^2-12288 i t^5+720 i t\nonumber\\
&&-6144 it^3 x^2+1152 i x^2 t-15360 t^4, \nonumber
\end{eqnarray}
$D_2$ is giving in the equation \eqref{D2}.
\begin{figure}[htb]
\centering
\subfigure[]{\includegraphics[height=35mm,width=35mm]{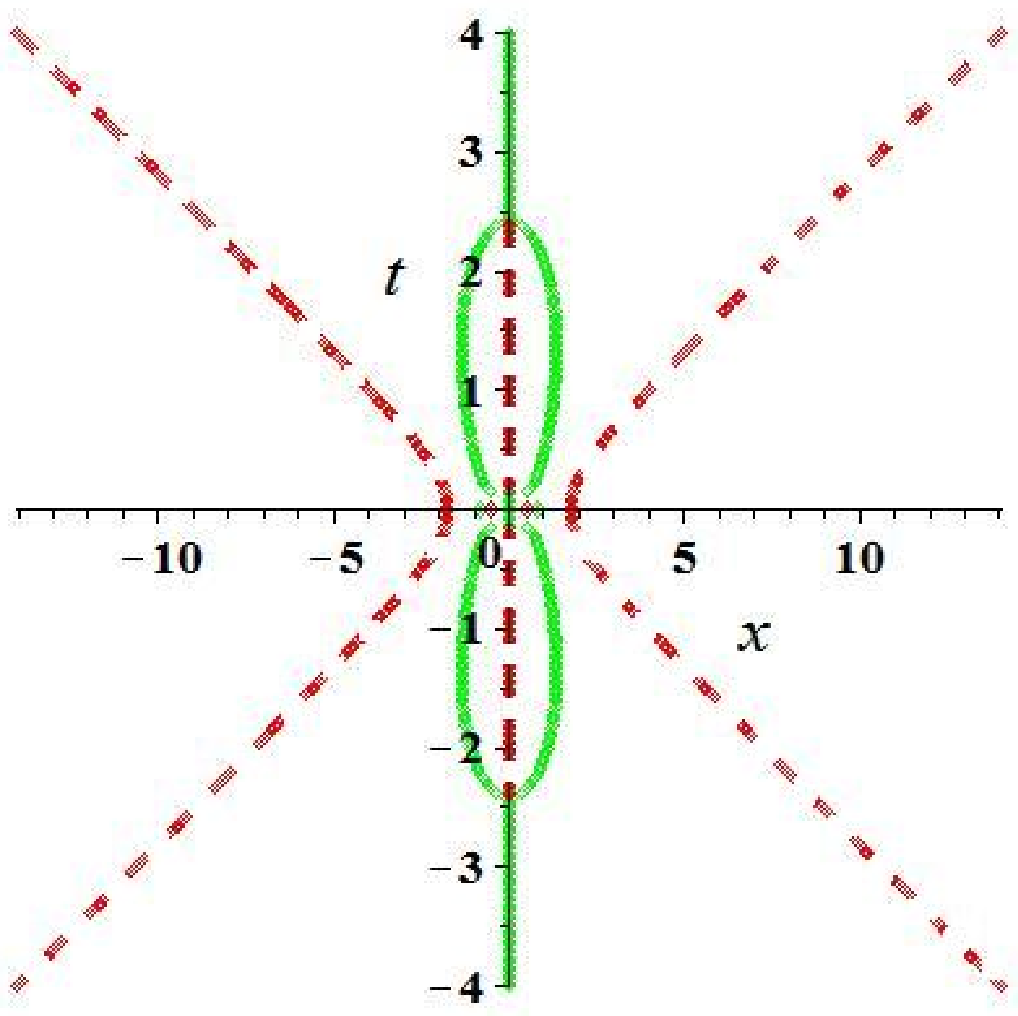}}
\hfil
\subfigure[]{\includegraphics[height=45mm,width=50mm]{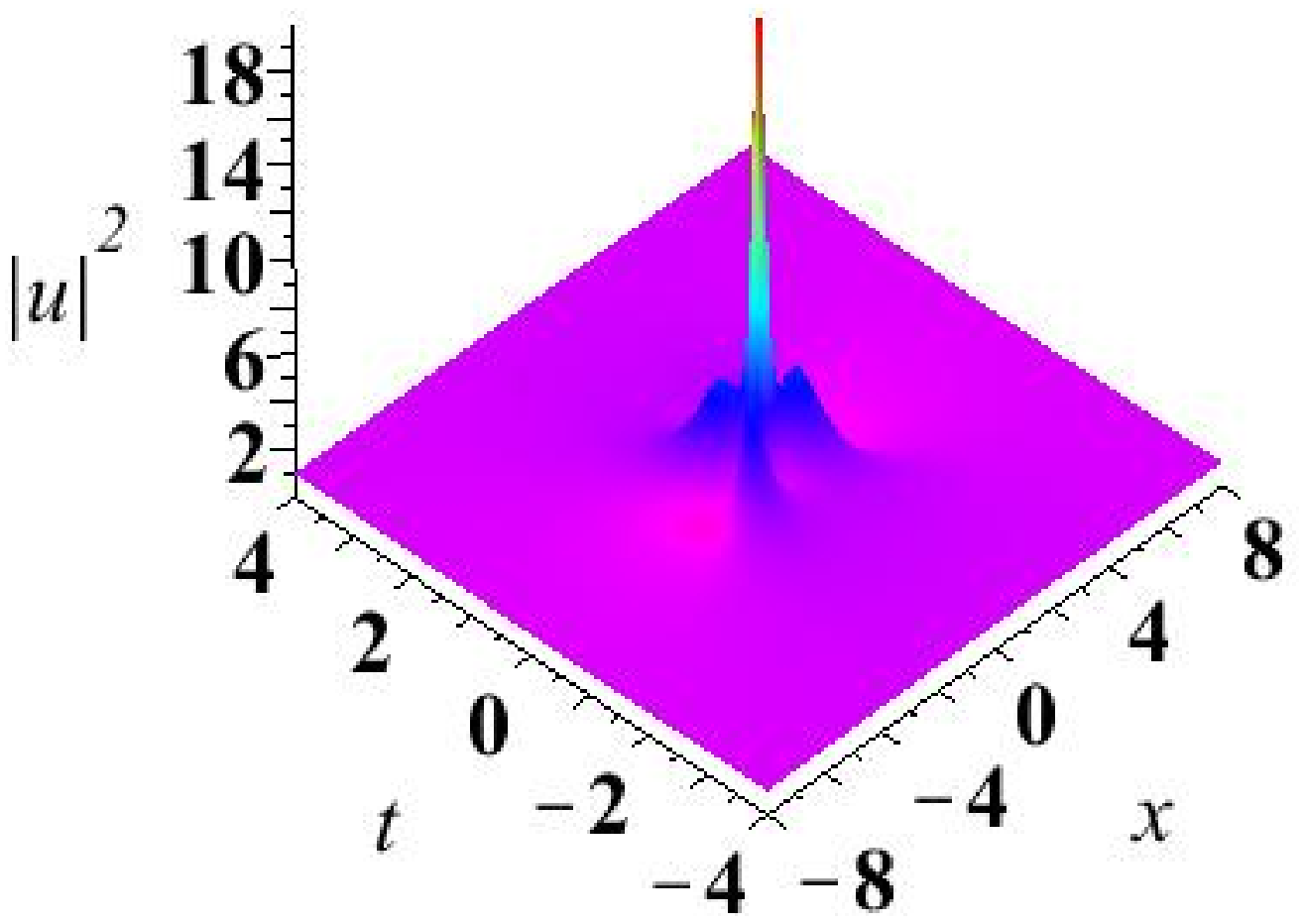}}
\hfil
\subfigure[]{\includegraphics[height=30mm,width=40mm]{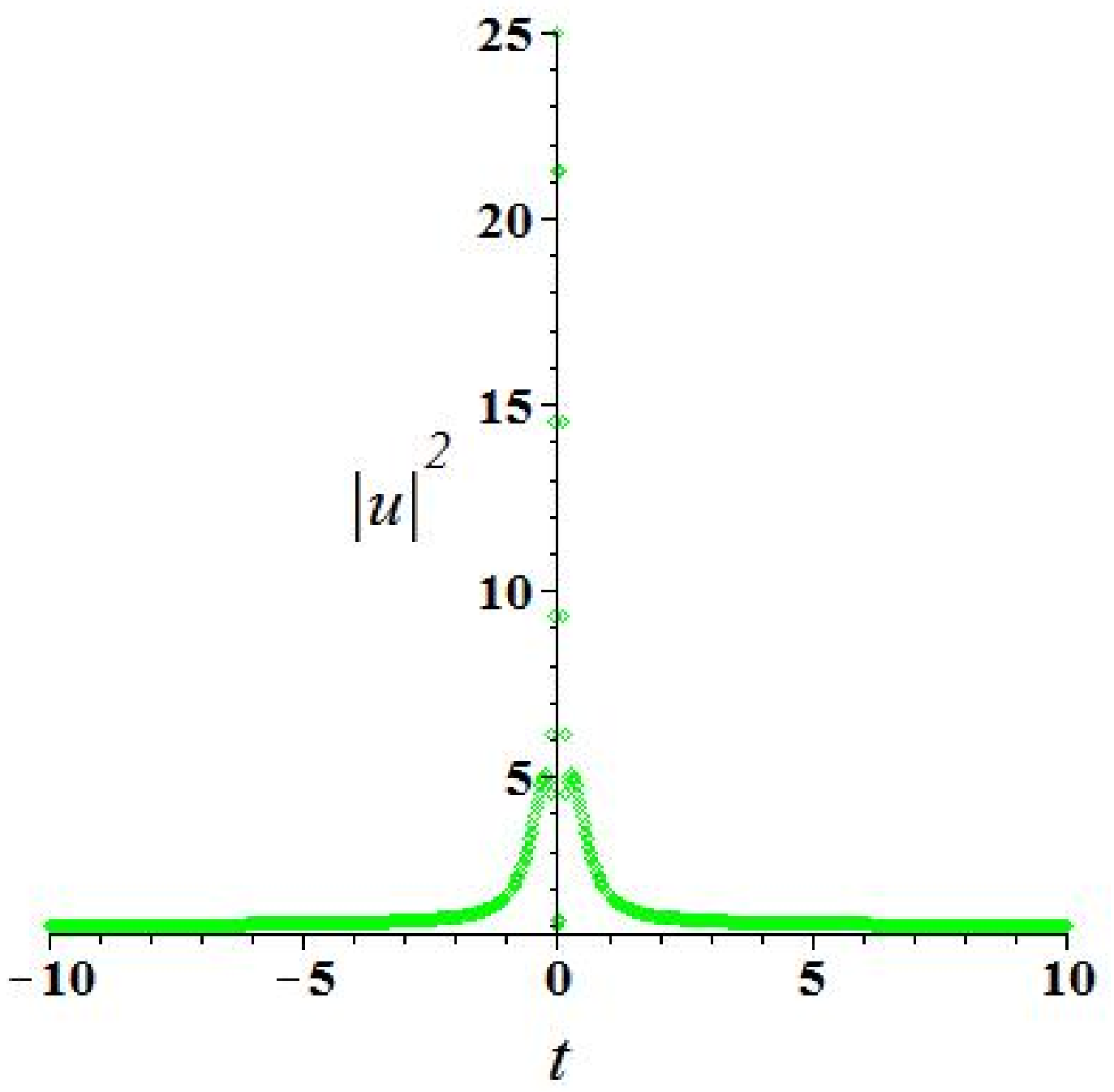}}
\hfil
\subfigure[]{\includegraphics[height=30mm,width=40mm]{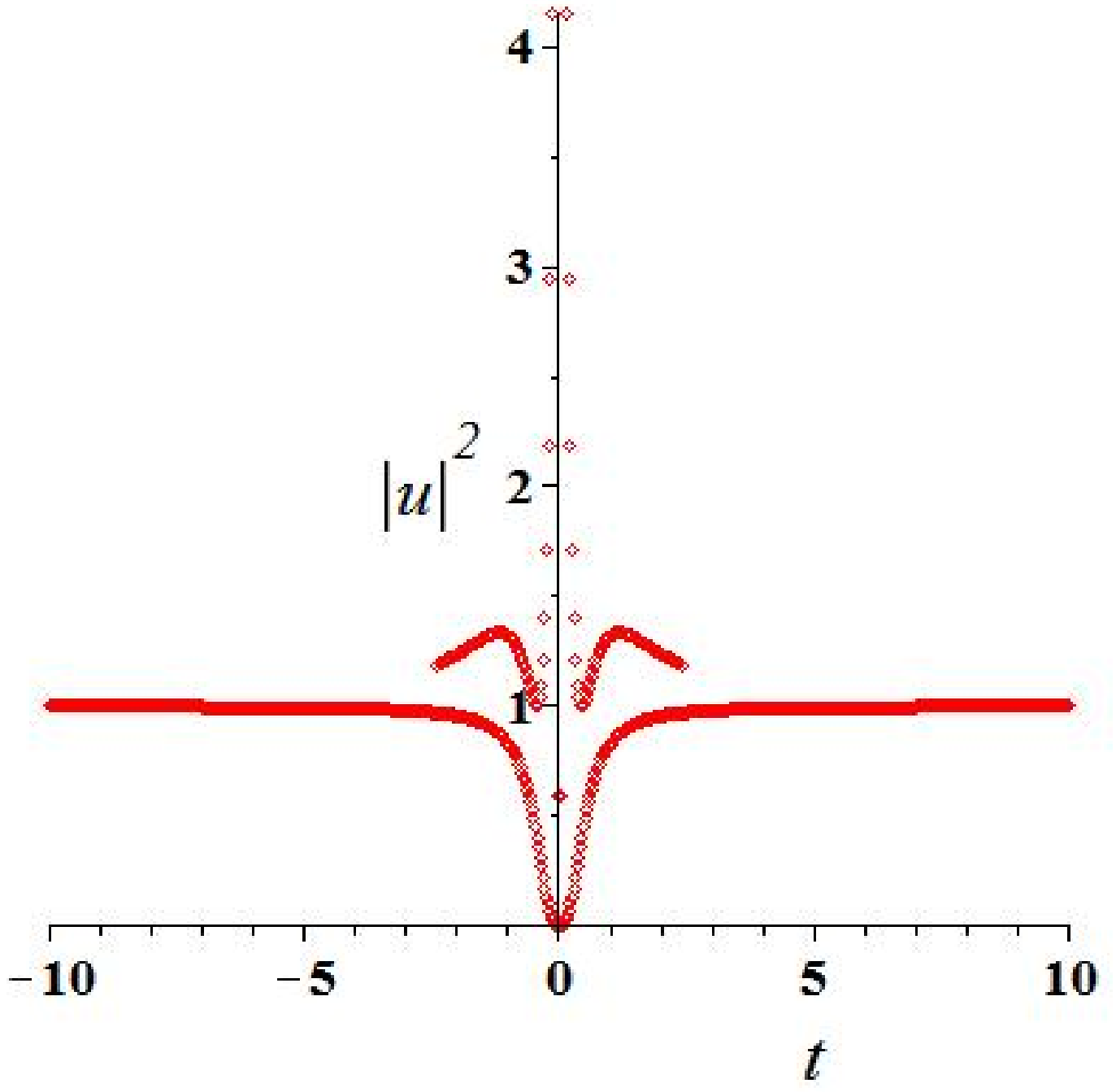}}
\caption{(color online) Standard 2-rd RW: (a) The trajectory of the
second-order RW which is symmetrical about the $x$ and $t$ axis. (b)
The density evolution of the second-order RW. (c) The value
evolution of RW's hump peak. (d) The value evolution of RW's
valleys. The parameters are $a=0$, and $b=0$. }\label{fig3}
\end{figure}

The trajectory of first order rogue wave can be derived exactly.
However, for the 2-nd rogue wave, we can not obtain exact expression
for the trajectory of 2-nd rogue wave, since the high order
algebraic equation emerge for these extreme points. When $x$ or $t$
$\rightarrow\infty$, we can readily prove that the trajectory of
2-nd order rogue wave is asymptotic to the first order rogue wave.
But we can not obtain the trajectory of 2-nd rogue wave in the
neighbourhood of $(x,t)=(0,0)$ with a simple way. To give the
trajectory of 2-nd rogue wave, we use the numerical method. We know
that the locations of humps and valleys can be used to describe RW's
trajectory. Thus we just need to obtain the trajectories of these
extreme points.

\begin{figure}[htb]
\centering
\subfigure[]{\includegraphics[height=35mm,width=35mm]{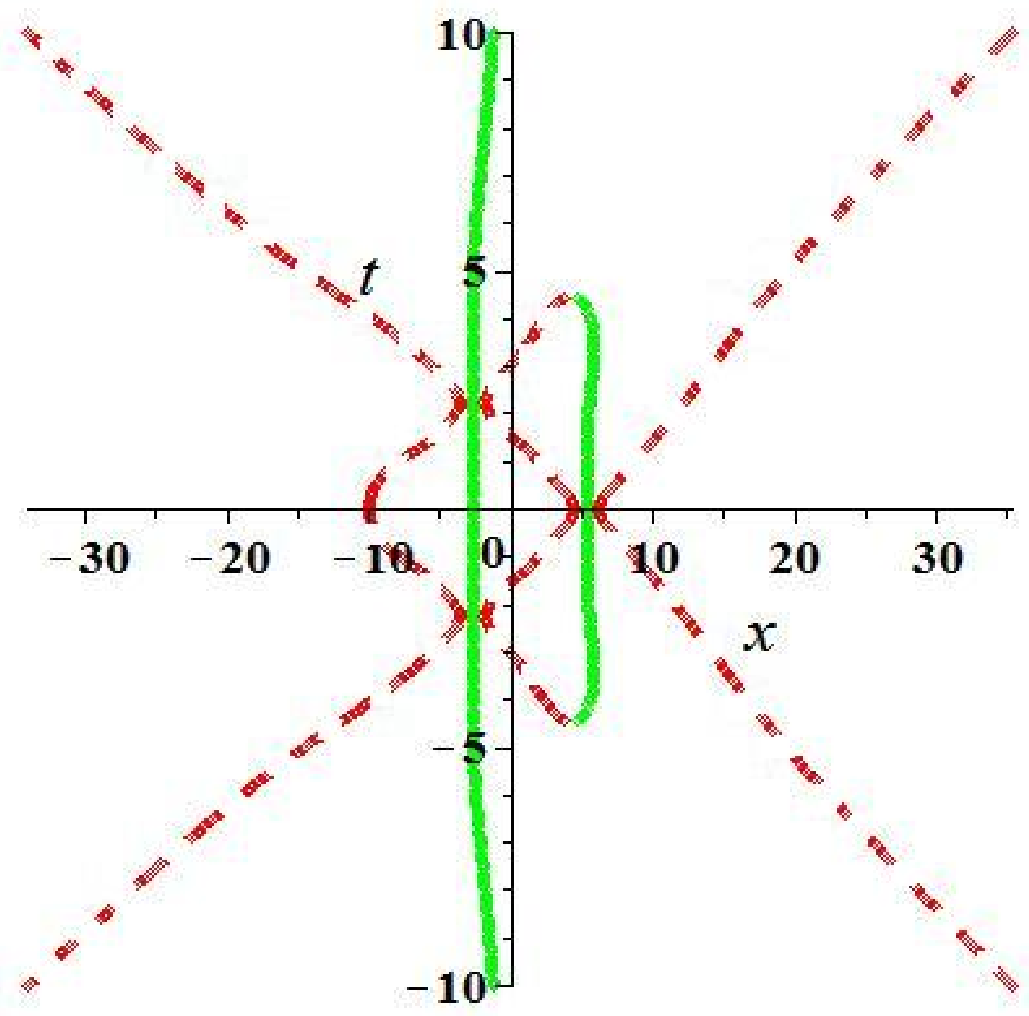}}
\hfil
\subfigure[]{\includegraphics[height=45mm,width=50mm]{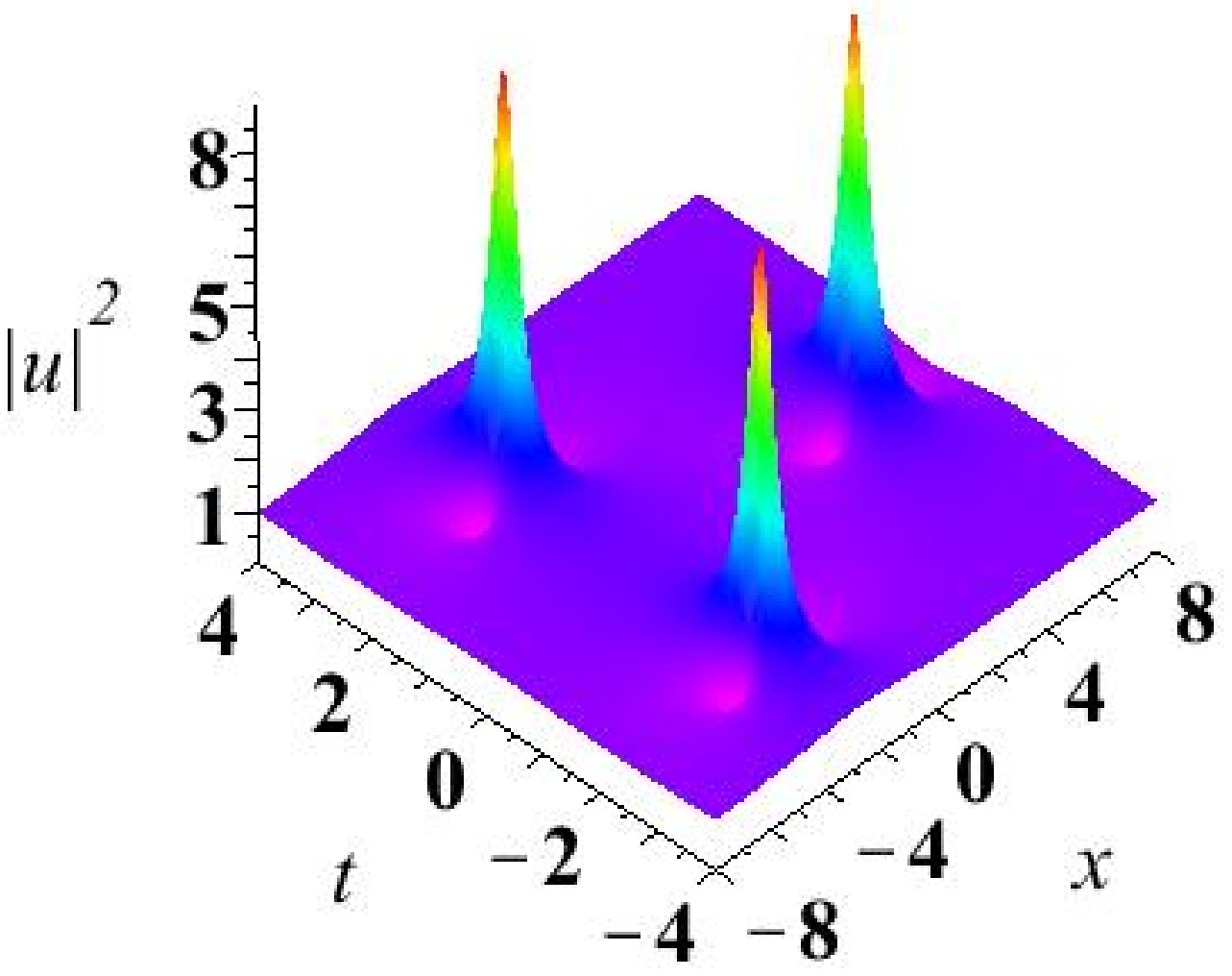}}
\hfil
\subfigure[]{\includegraphics[height=35mm,width=35mm]{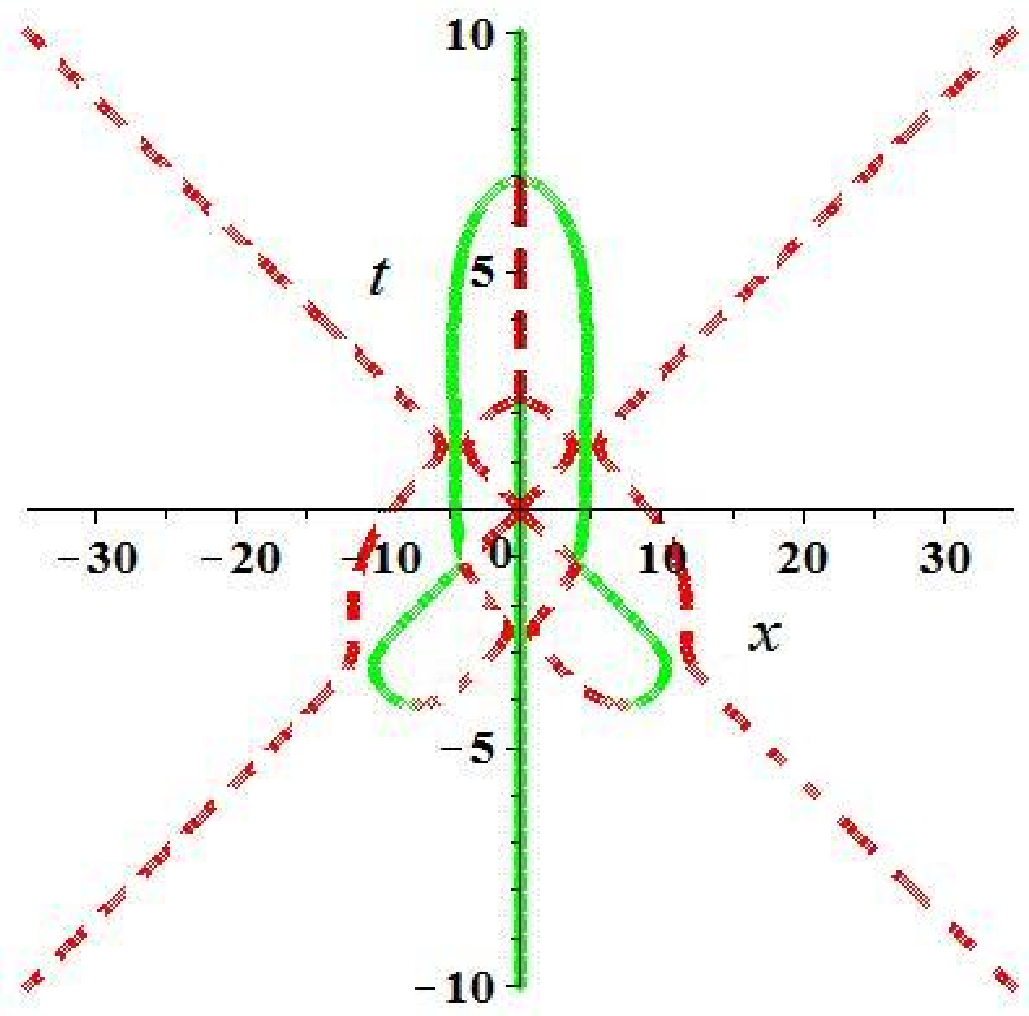}}
\hfil
\subfigure[]{\includegraphics[height=45mm,width=50mm]{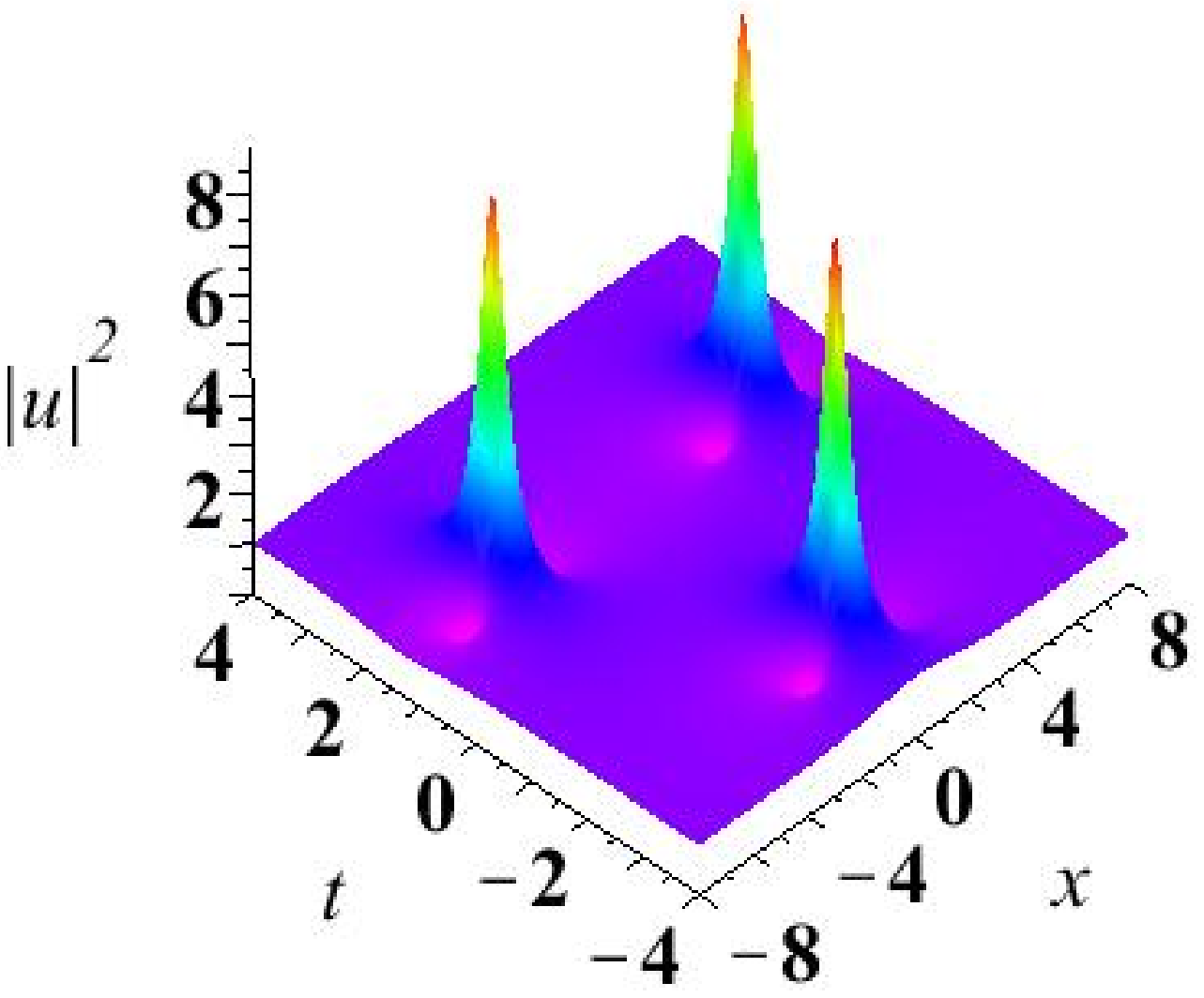}}
\caption{(color online) (a) The trajectory of the second-order RW
which is symmetrical about the $x$ axis with parameters $a=100$, and
$b=0$. (b) The density evolution of the second-order RW in (a). (c) The trajectory of the
second-order RW which is symmetrical about the $t$ axis with
parameters $a=0$, and $b=100$. (d) The density evolution of the
second-order RW in (c). }\label{fig4}
\end{figure}

We merely consider the three special cases for 2-nd rogue wave
solution. Firstly, we give the trace for the standard 2-nd rogue
wave solution in Fig. \ref{fig3} (a) and (b), which is symmetrical about both
the $x$ and $t$ axis. We use the numerical method to derive the peak
value curve and hole value curve for standard 2-nd rogue wave (Fig.
\ref{fig3} (c) and (d)). Compare the trajectory in Fig. \ref{fig3} with the one in
Fig. \ref{fig1}, we know that the the whole trajectories of them are similar,
but the trajectories and structures are distinctive from each other
near the location where highest peak emerges. Compare the curves in
Fig. \ref{fig3} (c) and (d) with Fig. \ref{fig2}, we can see that the two curves are
very similar. But they possess different peak values. The maximum of
peak value curve for 2-nd rogue wave is 25. But the maximum of hump
value curve for first order rogue wave is 9. The valley value of 2-nd
rogue wave is also similar with first order rogue wave. But the valley curve near $t=0$  is higher than the
background, and the maximum of the curve is about $4.2$.

\begin{figure}[htb]
\centering
\subfigure[]{\includegraphics[height=35mm,width=35mm]{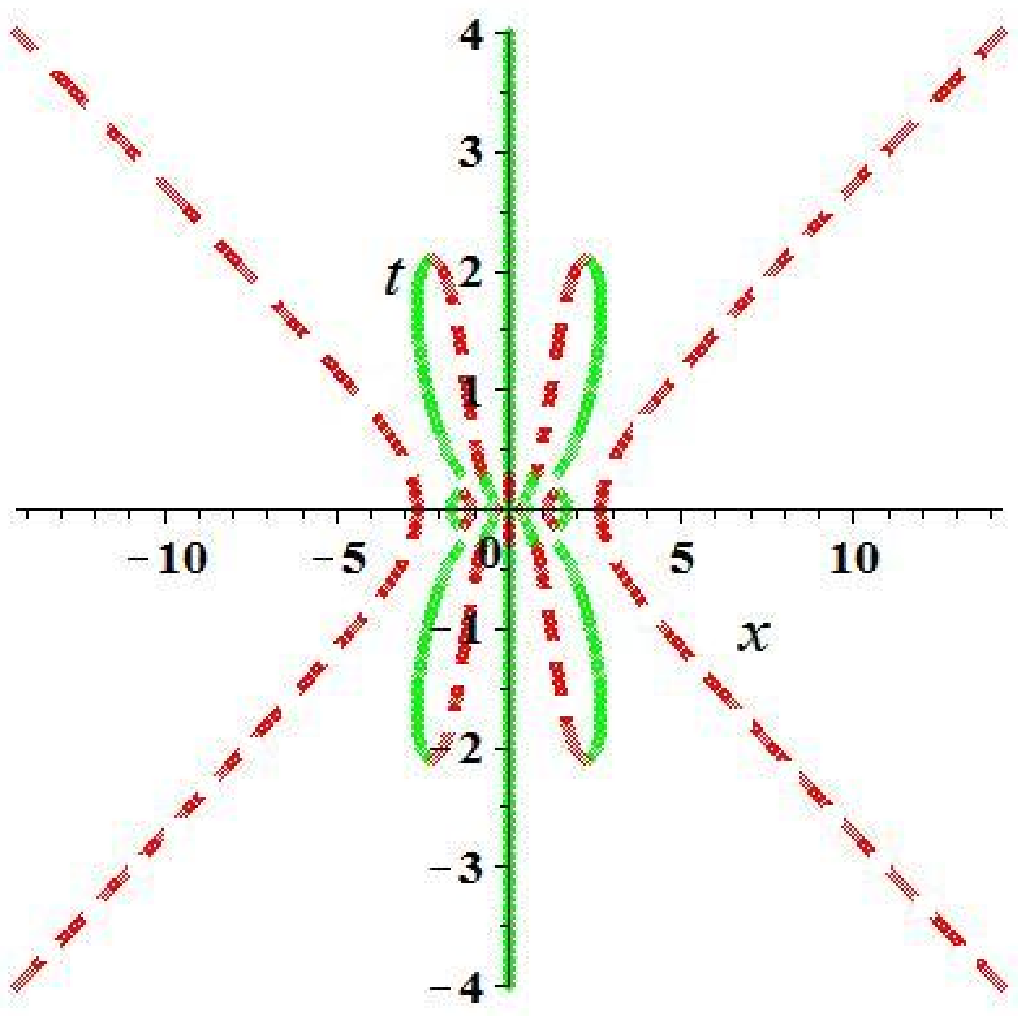}}
\hfil
\subfigure[]{\includegraphics[height=45mm,width=50mm]{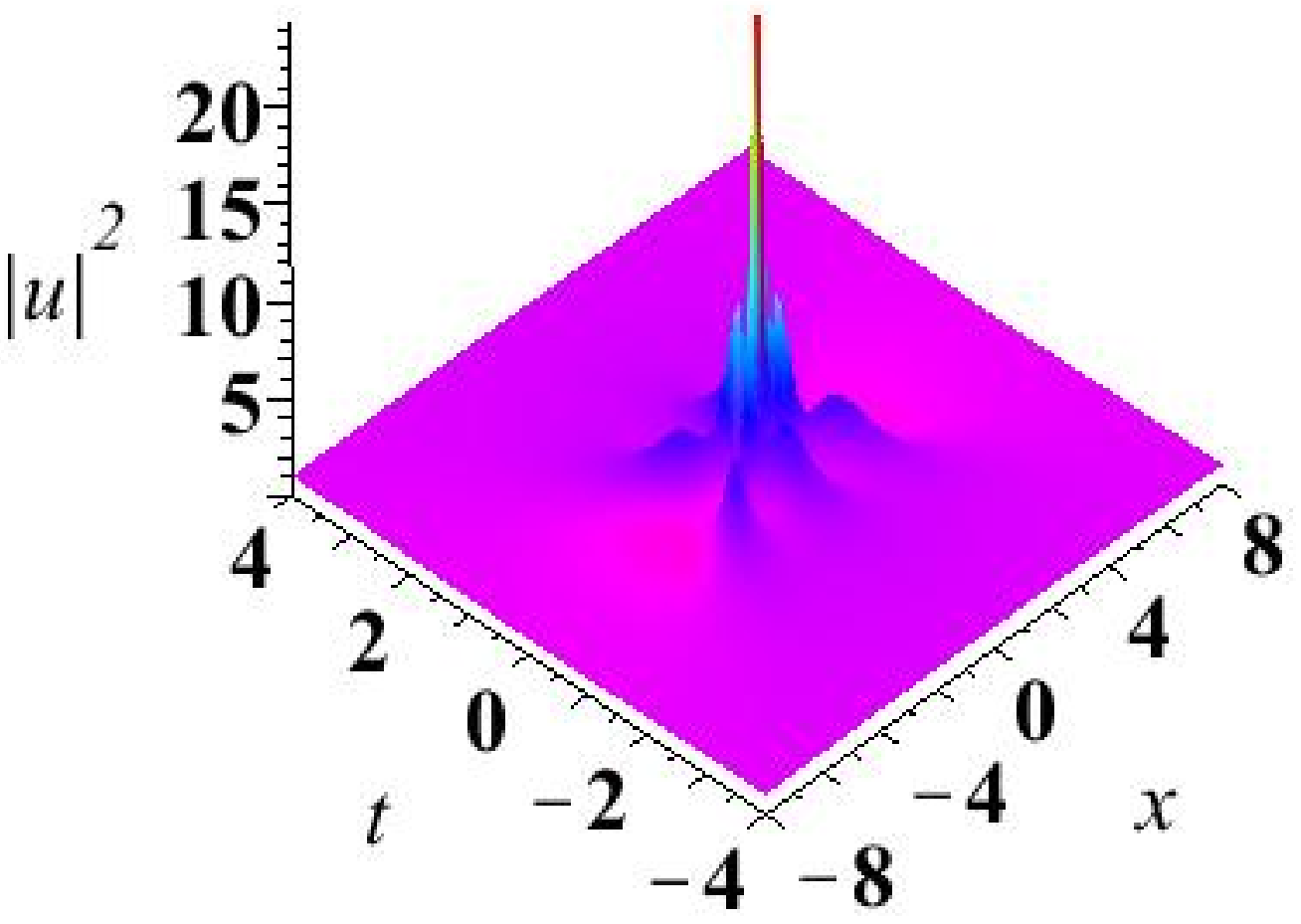}}
\hfil
\subfigure[]{\includegraphics[height=35mm,width=35mm]{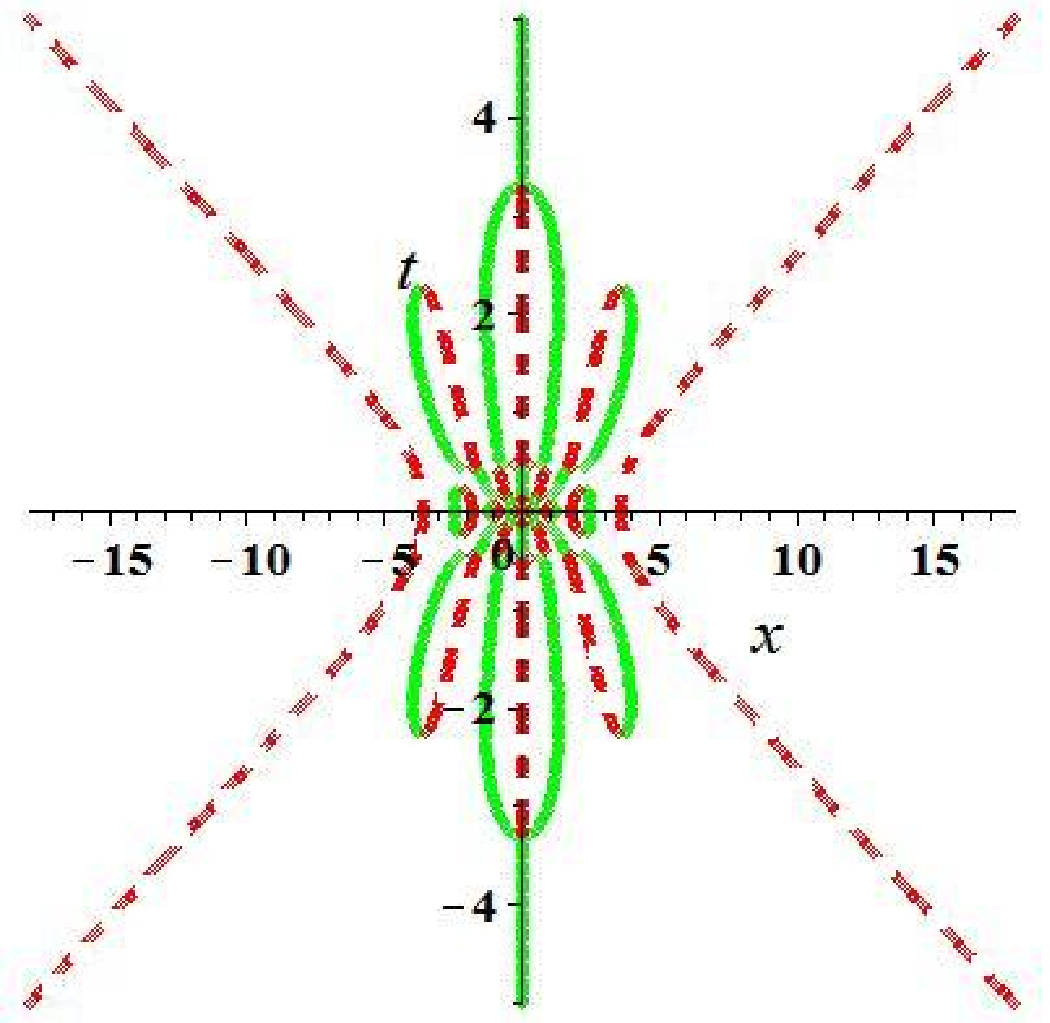}}
\hfil
\subfigure[]{\includegraphics[height=45mm,width=50mm]{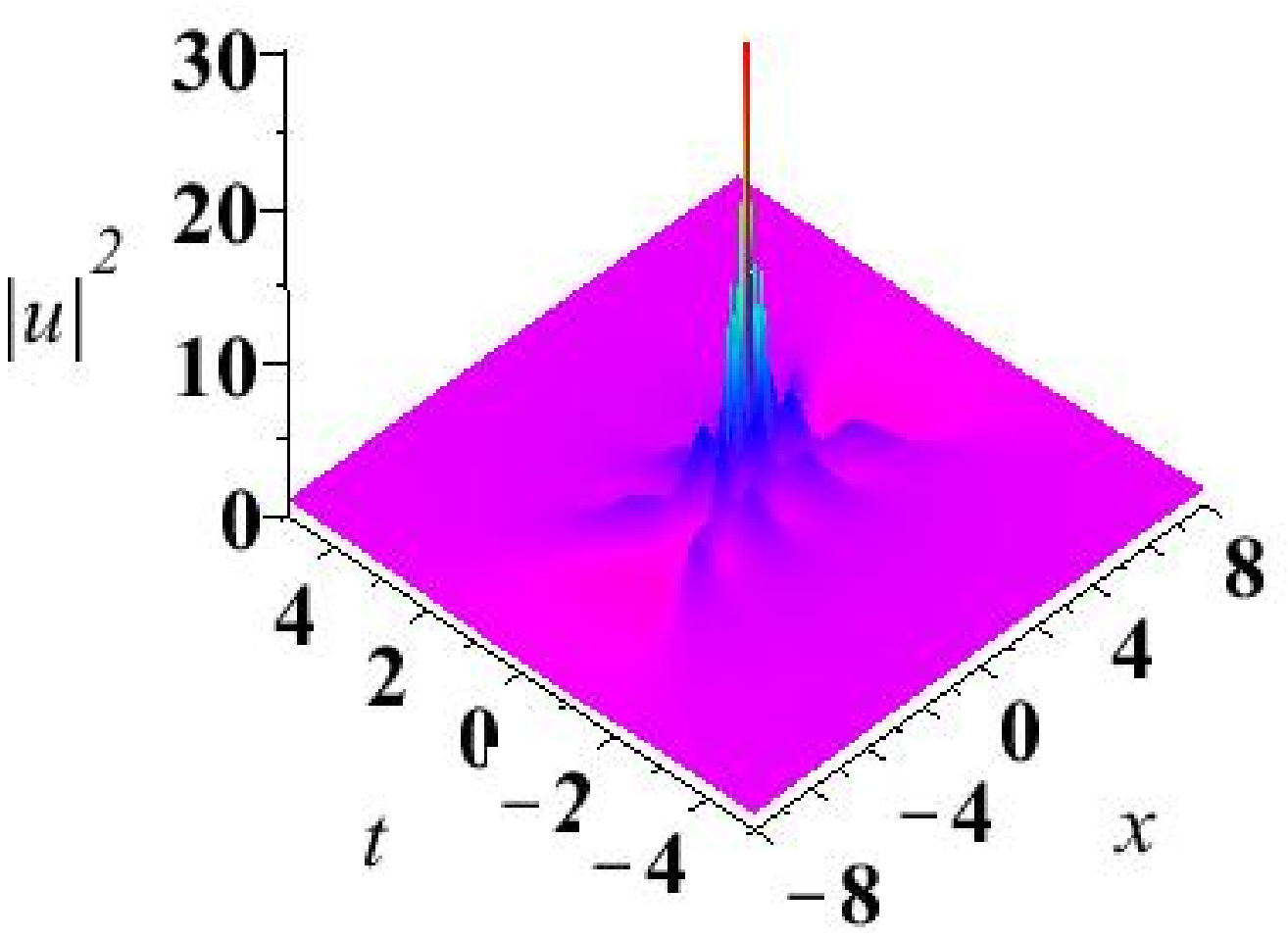}}
\caption{(color online) (a) The trajectory of the third-order RW
which is symmetrical about the $x$ and $t$ axis. (b) The density
evolution of the third-order RW. (c) The trajectory of the
fourth-order RW which is symmetrical about the $x$ and $t$ axis. (d)
The density evolution of the fourth-order RW.}\label{fig5}
\end{figure}

For second-order RW, there can be three symmetry RWs in the
temporal-spatial distribution \cite{Akhmediev3}. As examples, we
give other two special cases for the trajectory of 2-nd rogue wave
solution in Fig. \ref{fig4}, which are symmetrical about the $x$ and
$t$ axis separately. From the pictures, we can see that the
trajectory of RW is consistent with the fundamental RW. Each center
of the ``X" shape corresponds to the highest peak's location. This
can be verified by comparing the locations in Fig. \ref{fig4}.

\section{The classification of high-order order rogue wave by parameters}

In this subsection, we give the classification of high-order rogue
wave through parameters $s_i$. For instance, there is no parameters $s_i$ in the standard first order RW. So there is only one kind.
The general second order RW has one parameter $s_1$, we classify it by whether or not the parameter $s_1$ equals to zero.
If parameter $s_1=0$, we denote it as the type $[0]$. Otherwise, we denote it as the type $[1]$. The general third-order RW
has two parameters $s_1$, $s_2$. We still classify it by whether or not the parameters $s_1$, $s_2$ equal to zero. If $s_1=s_2=0$, this is the standard case, we denote it
as the type $[0,0]$. The other cases are type $[0,1]$, $[1,0]$ and $[1,1]$. Similarly, the general fourth order RW has three parameters $s_1$, $s_2$ and $s_3$.
Then the types are $[0,0,0]$, $[1,0,0]$, $[0,1,0]$, $[0,0,1]$, $[1,1,0]$, $[1,0,1]$, $[0,1,1]$ and $[1,1,1]$.

\begin{figure}[htb]
\centering
\subfigure[]{\includegraphics[height=42mm,width=50mm]{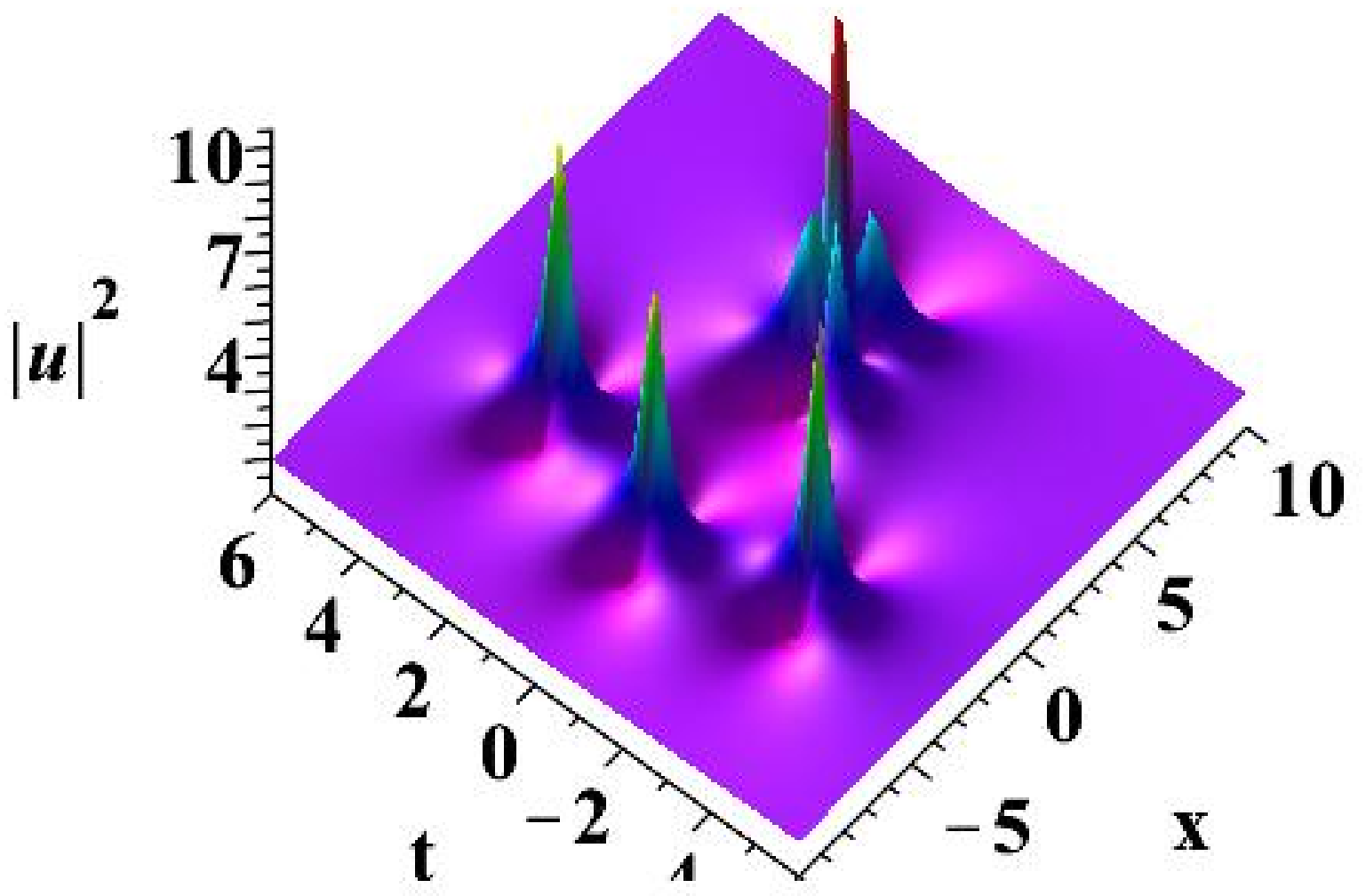}}
\hfil
\subfigure[]{\includegraphics[height=42mm,width=50mm]{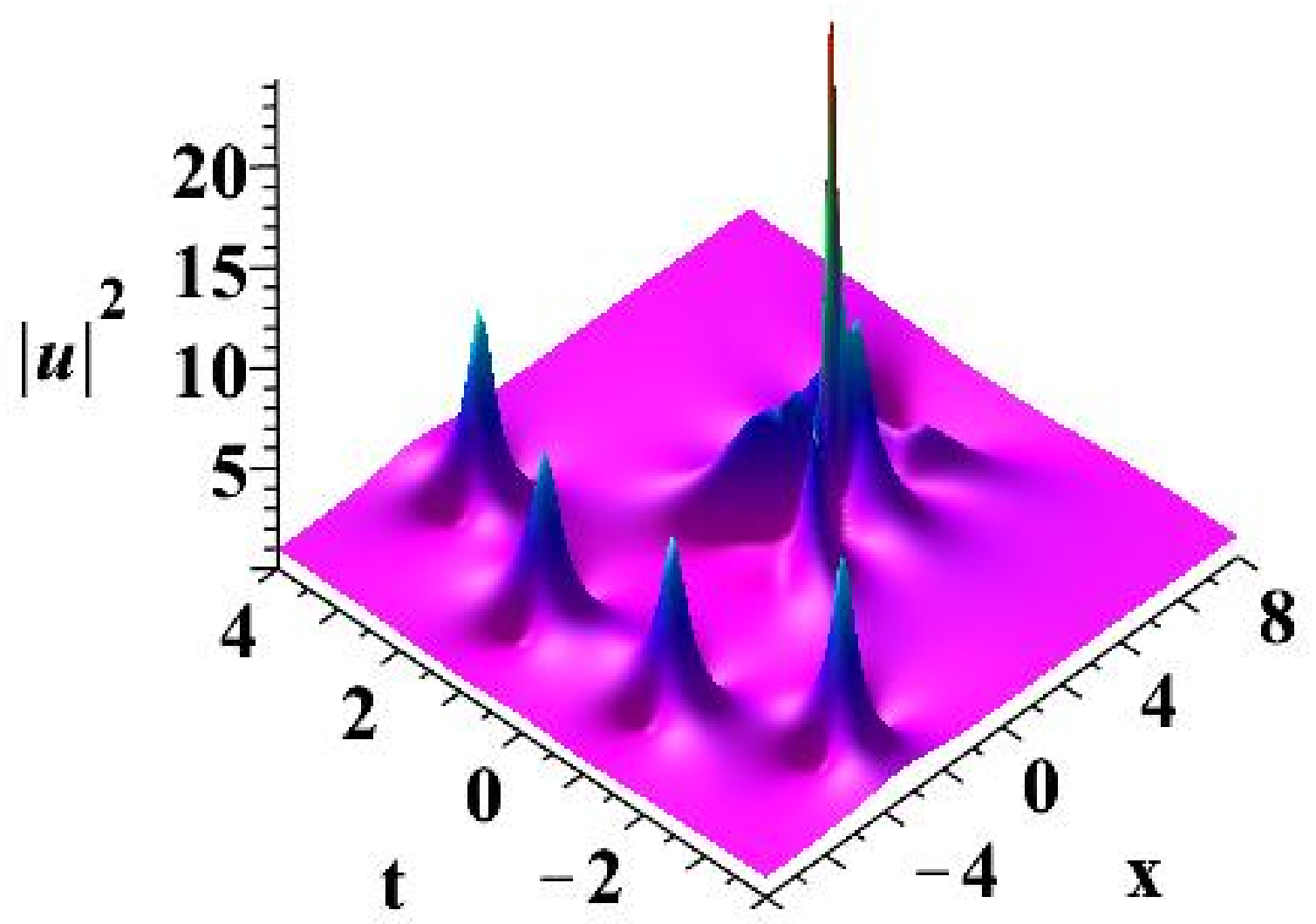}}
\caption{(color online) (a). The ``claw-like" structure of the third-order rogue wave with $s_1=31.2$, $s_2=500$.
 On the left side, there are three RWs form an arc. On the right side, there is a second- order RW with the highest peak.
 (b) The standard ``claw" structure of the fourth-order RW with $s_1=12$, $s_2=101$, $s_3=1020$. On the left side,
 there are four RWs forming an arc. On the right side, there is a third-order RW with the highest peak}\label{fig6}
\end{figure}

Indeed, there are lots of interesting spatial-temporal distribution pictures hiding in our classification. For
instance, in the general second-order RW, the type $[0]$ corresponds the standard one and the type $[1]$ corresponds the ``triplets"
\cite{Akhmediev3}. For the general third-order rogue wave, the type $[0,0]$ corresponds the ``standard" structure \cite{N.A},
the type $[1,0]$ corresponds the ``triangular cascades" structure \cite{KAA3},
the type $[0,1]$ corresponds the ``pentagram" structure \cite{KAA3}. The ``claw" structure\cite{KAA2}
and the ``arrow" structure \cite{KAA2} are all involved in the type $[1,1]$.
The ``claw" structure was first derived in reference \cite{KAA2} by numeric method. Different from them,
we also present the ``claw" structure of the third-order RW by explicit analytical expression with proper ratio to parameters $s_1$ and $s_2$ (Fig. \ref{fig6} a). But we fail to find the explicit ratio value and just report its existence.

\begin{figure}[htb]
\centering
\subfigure[]{\includegraphics[height=42mm,width=50mm]{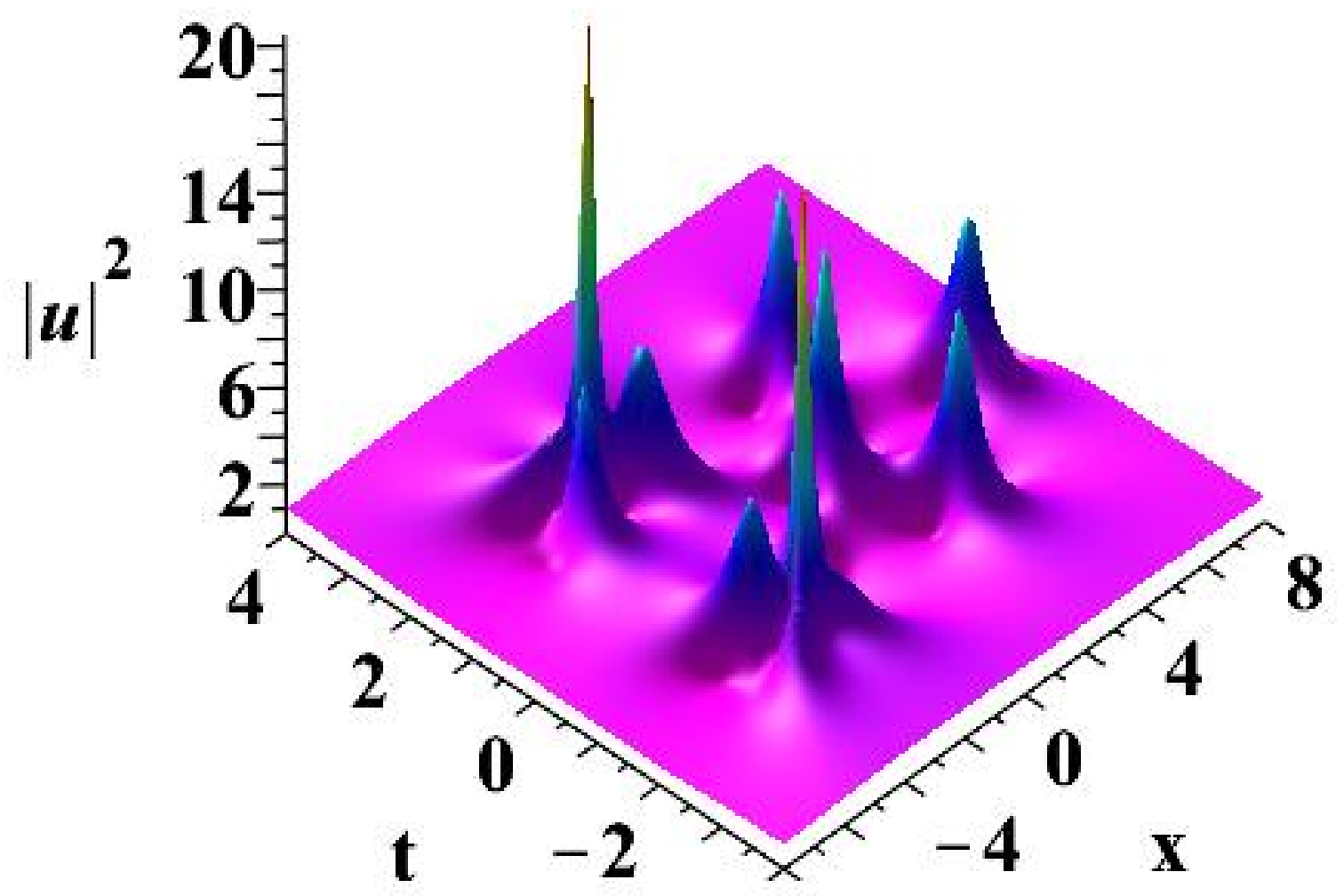}}
\hfil
\subfigure[]{\includegraphics[height=42mm,width=50mm]{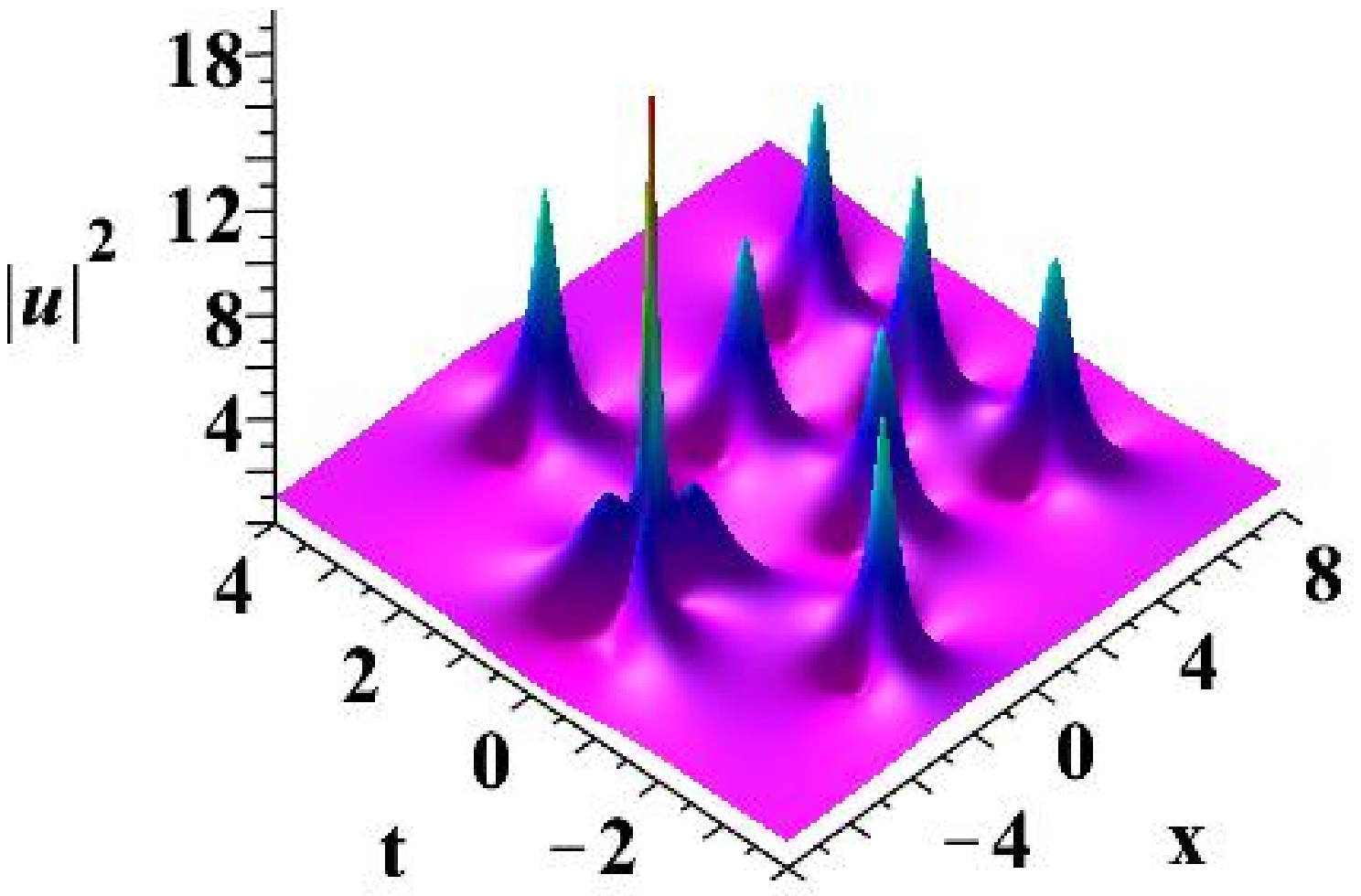}}
\caption{(color online) (a) The ``double column" structure of the
fourth-order rogue wave with $s_1=9$, $s_2=0$, $s_3=453$. There are
two second-order RWs on the left side look like two columns. On the
right side, there are four 1st-order RWs arranging with a
quadrangle. (b) The ``claw-line-I" structure of the fourth-order RW
with $s_2=104$, $s_3=940$, $s_1=0$. On the left hand, there is a
claw-like structure. And the number of claw is 4. Besides this,
there are three RWs arrange a line. Thus we call it as
``claw-line-I" type.}\label{fig7}
\end{figure}

\begin{figure}[htb]
\centering
\subfigure[]{\includegraphics[height=42mm,width=50mm]{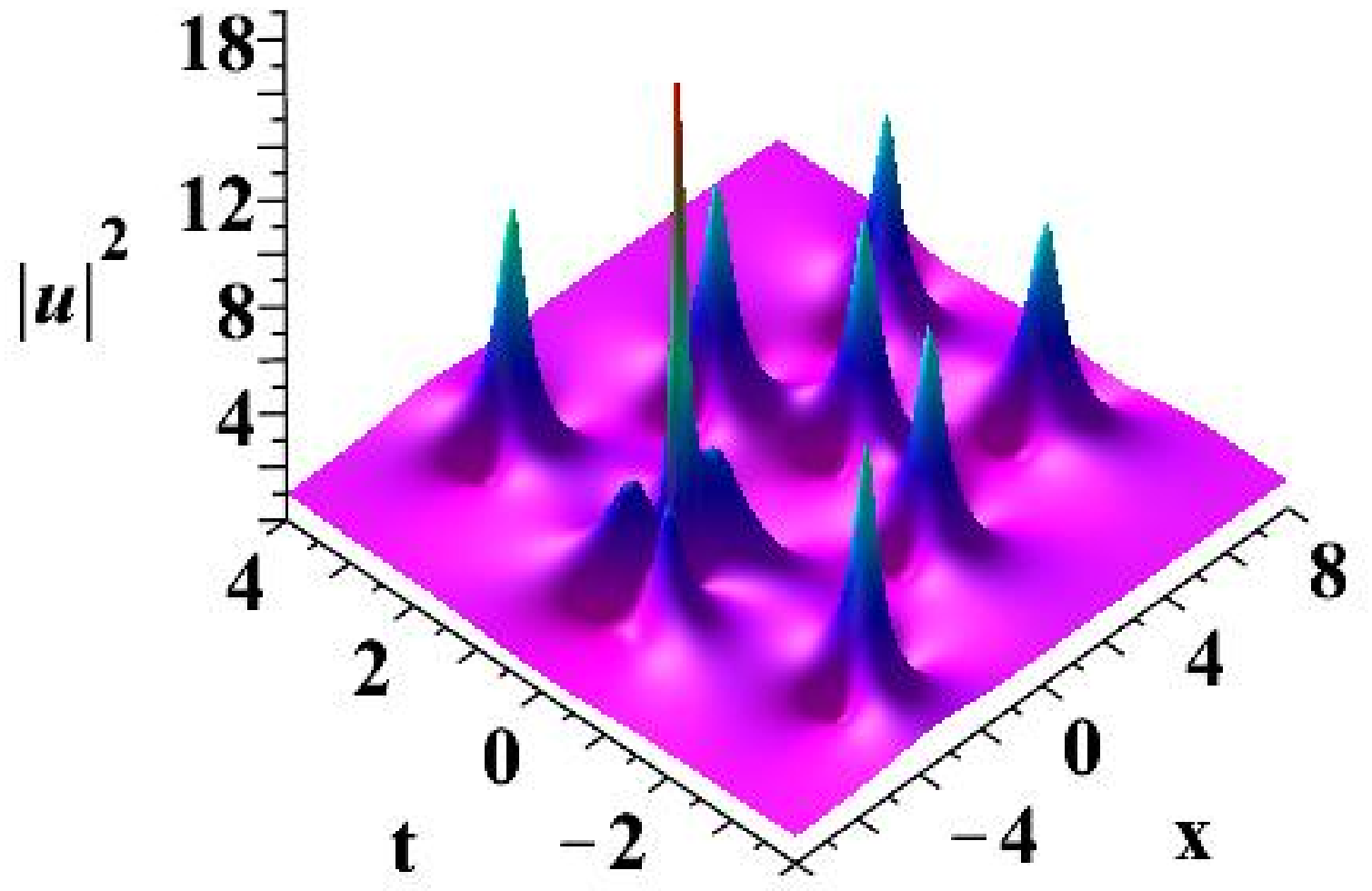}}
\hfil
\subfigure[]{\includegraphics[height=42mm,width=50mm]{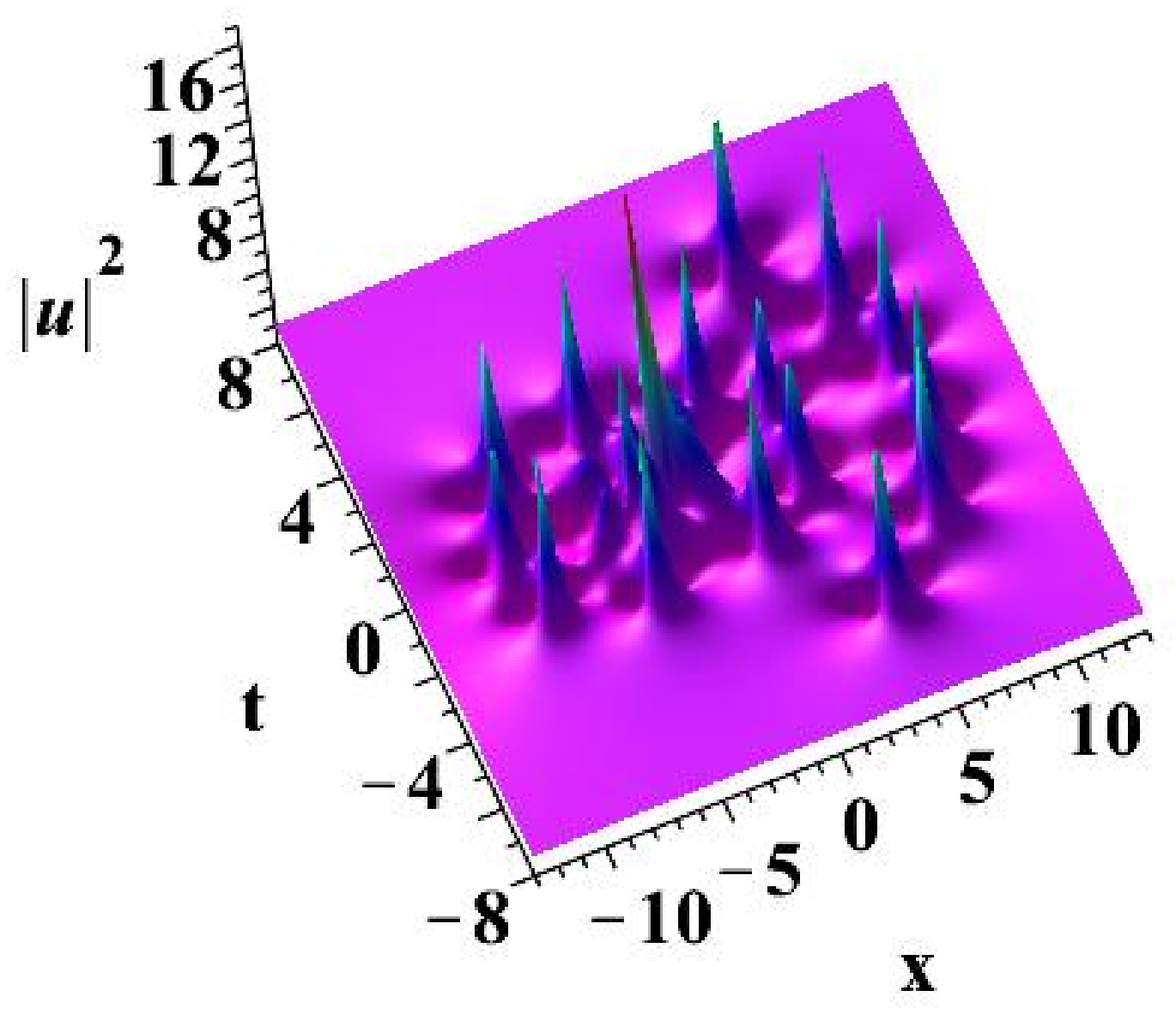}}
\caption{(color online) (a) The ``claw-line-II" structure of the
fourth-order RW with $s_1=8$, $s_2=108$, $s_3=0$. On the left hand,
there is clawlike structure. And the number of claw is 5. Besides,
there are two RWs arrange a line on the right hand. Thus we call it
as ``claw-line-II" type. (b) The ``circle-arc" structure of the
sixth-order RW with $s_1=-12$, $s_2=-100$, $s_3=-1000$,
$s_4=-10800$, $s_5=-150000$. There is a enneagram on the left hand.
The rest six fundamental RWs arrange with an arc.}\label{fig8}
\end{figure}

The general fourth order RW is obtained by above formula \eqref{NLS-formula}.
In previous research \cite{Ling}, we know that the type $[0,0,0]$ corresponds the ``standard" structure \cite{N.A}, type $[1,0,0]$ corresponds ``triangle" structure \cite{KAA3},
type $[0,1,0]$ corresponds ``pentagram" structure \cite{KAA3}, type $[0,0,1]$ corresponds ``heptagram" structure \cite{KAA3}. In what following, we look for some interesting structure in the other four types.
Firstly, we look for interesting structure in the type $[1,1,1]$. Choosing the parameters $s_1$, $s_2$
and $s_3$ with proper proportion, we can obtain the standard ``\textbf{claw-like}" RW (Fig. \ref{fig6} (b)), which possesses
one 3rd-order RW and four 1st-order RWs structure. This structure has been obtained in reference
\cite{KAA2} by numeric method. Here we also obtain it by the explicit rational expression.
Secondly, we search the interesting structures in the type $[1,0,1]$. By choosing the parameters $s_1$ and $s_3$ with proper ratio and $s_2=0$, we can obtain a new kind of structure ``\textbf{double column}" structure (Fig. \ref{fig7} (a)),
which possesses two standard 2nd-order RWs and four first-order
RWs. Finally, we obtain two new types of claw-like RW structure in the type $[0,1,1]$ and $[1,1,0]$ respectively.
The first type ``\textbf{claw-line-I} " structure type is shown in Fig. \ref{fig7} (b). This type of RW can be obtained
by setting parameters $s_2$ and $s_3$ with appropriate ratio and $s_1=0$.
This RW possesses a standard 2rd-order RW and seven 1st-order RWs, for which the four fundamental RWs arrange with an arc, and the other three fundamental RWs arrange with a line.
The second type ``\textbf{claw-line-II}" RW is shown in in Fig. \ref{fig8} (a), which
possesses the similar structure with one 2rd RWs and seven 1st RWs,
which can be obtain by choosing $s_1$ and $s_2$ with proper ratio and $s_3=0$.
Among them, the five fundamental RWs arrange with an arc, the other two fundamental RWs are located in the outer edge of the arc. Indeed, there maybe exist other interesting temporal-spatial distribution pictures \cite{Gaillard1}. Here we merely list some
interesting structures.

For general high-order rogue wave solution, we have the following
classification. The general N-th order rogue wave solution possesses
$N-1$ parameters. So it possesses $2^{N-1}$ cases. As the order
increase, the types of structure become more and more abundant. In
this paper, we don't research it in detail. As an example, we show
the sixth-order RW with ``\textbf{circle-arc}" structure
(Fig \ref{fig8} (b)), which is consist of an ``enneagram" with a
circle boundary and an ``arc" with six fundamental RWs arranged. To our knowledge, this is a new structure for sixth-order RW, which would enrich our knowledge about complex localized waves in the related physical systems.

\section{discussion and conclusion}
We propose a simple representation for generalized RW solution,
which can be used to get arbitrary order RW solution and observe its
dynamics conveniently. Based on the solution, we investigate the
trajectories of them though defining the properties function. We
find that the fundamental RW's valleys have  ``X" shaped
trajectories. For higher-order RW, the whole trajectory is similar
to the fundamental one's. But they are different near the moment
when the highest peak emerges, such as the trajectories of the
third-order and fourth-order RW in Fig. \ref{fig5}.  Finally, the
classification of general high-order RW solution is discussed. We
find some new structures for high-order RWs by choosing different
parameters with some proper ratios, such as ``double column"
structure, ``claw-line" structure, and ``circle-arc" structure. But
we fail to obtain the exact value of the ratio for these new
structure RWs. The systemic classification on them should be done in
the near future.

\section*{Appendix: The explicit expressions of $F_j$}

\begin{eqnarray*}
F_0&=&36+4608 ac+18432 {a}^{2}{b}^{2}+2304 {d}^{2}+2304 {c}^{2}+9216 {a}^{4}+9216 {b}^{4}+3456 {b}^{2}+3456 {a}^{2}+4608 bd \nonumber\\
     &&+ \left( 18432 b+36864 {a}^{2}d-36864 {b}^{2}d+11520\,d-73728 abc+73728 {a}^{2}b+73728 {b}^{3} \right) t\nonumber\\
     &&+ \left( 129024 {a}^{2}+26496+36864 {c}^{2
}+276480 {b}^{2}+36864 {d}^{2}-110592 ac-110592 bd \right) {t}^{2}\nonumber\\
     &&+ \left( -98304 {b}^{3}-98304 {a}^{2}b-36864 d+620544 b \right) {t
}^{3}\nonumber\\
     &&+ \left( 196608 ac+49152 {a}^{2}-540672 {b}^{2}+196608  bd+
654336 \right) {t}^{4}\nonumber\\
     &&+ \left( {\frac {3342336}{5}} d-{\frac {2064384
}{5}} b \right) {t}^{5}+ \left( {\frac {1835008}{5}} {a}^{2}+{\frac
{786432}{5}} {b}^{2}+1114112 \right) {t}^{6}\nonumber\\
     &&+ \left( {\frac {8912896}
{5}} b+{\frac {1048576}{5}} d \right) {t}^{7}+{\frac {21757952}{5}}
 {t}^{8}+{\frac {8388608}{15}} b   {t}^{9}\nonumber\\
     &&+{\frac {58720256}{25}} {t}^
{10}+{\frac {67108864}{225}} {t}^{12}, \nonumber
\end{eqnarray*}
\begin{eqnarray*}
F_1&=&18432 {b}^{2}c-36864 abd+2304 a-18432 {a}^{2}c+1152 c+ \left(
110592 bc-110592 ad \right) t\nonumber\\
     &&+ \left( -147456 a{b}^{2}-147456 {a}^
{3}+202752 c+82944 a \right) {t}^{2}\nonumber\\
     &&+ \left( -196608 ad+196608 bc-
983040 ab \right) {t}^{3}+ \left( 884736 c-737280 a \right) {t}^{4}
-{\frac {3145728}{5}} ab{t}^{5}\nonumber\\
     &&+ \left( 524288 c-{\frac {262144}{5}}
 a \right) {t}^{6}+{\frac {4194304}{5}} a{t}^{8},
\end{eqnarray*}
\begin{eqnarray*}
F_2&=&-9216 ac+864+9216 {c}^{2}+9216 {d}^{2}-9216 bd -4608 {a}^{2}-4608
 {b}^{2}+ \left( -41472 b+73728 {b}^{3}-46080 d+73728 {a}^{2}b
 \right) t\nonumber\\
     &&+ \left( 663552 {b}^{2}+73728 {a}^{2}-41472 \right) {t}^{2
}+ \left( -294912 d+1622016 b \right) {t}^{3}\nonumber\\
     &&+ \left( 1425408+589824
 {b}^{2}-196608 {a}^{2} \right) {t}^{4}+ \left( {\frac {12976128}{5}
} b-{\frac {2359296}{5}} d \right) {t}^{5}\nonumber\\
     &&+3932160 {t}^{6}+{\frac {
6291456}{5}} {t}^{8}+\frac {33554432}{75} {t}^{10},
\end{eqnarray*}
\begin{eqnarray*}
F_3&=&-4608 c+12288 {a}^{3}-3840 a+12288 a{b}^{2}+ \left( 49152 b c+
147456 a b-49152 a d \right) t\nonumber\\
     &&+ \left( 270336 a+49152 c \right) {t}^
{2}+524288 a b{t}^{3}+ \left( -131072 c+1638400 a \right) {t}^{4}+{
\frac {8388608}{15}} a{t}^{6},
\end{eqnarray*}
\begin{eqnarray*}
F_4&=&3072 {b}^{2}+15360 {a}^{2}-12288\,ac-12288 bd+960+ \left( 18432 b-
36864 d \right) t+ \left( 147456 {a}^{2}-49152 {b}^{2}+61440
\right) {t}^{2}\nonumber\\
&&+ \left( -65536\,d-294912\,b \right) {t}^{3}-98304 {t
}^{4}-{\frac {1048576}{5}} b{t}^{5}+{\frac {1048576}{15}} {t}^{6}+{
\frac {4194304}{15}}\,{t}^{8},
\end{eqnarray*}
\begin{eqnarray*}
F_5&=&-{\frac {18432}{5}} c+{\frac {27648}{5}} a-{\frac {196608}{5}} bat+
 \left( -{\frac {294912}{5}} c-{\frac {49152}{5}} a \right) {t}^{2}+
{\frac {524288}{5}} a{t}^{4},\nonumber\\
F_6&=&{\frac {12288}{5}} {a}^{2}+{\frac {13312}{5}}+{\frac {28672}{5}} {b}
^{2}+ \left( 16384 d+{\frac {8192}{5}} b \right) t+{\frac {49152}{5}
} {t}^{2}-{\frac {1048576}{15}} b{t}^{3}-{\frac {262144}{5}} {t}^{4
}+{\frac {4194304}{45}} {t}^{6},\nonumber\\
F_7&=&{\frac {8192}{5}} c-{\frac {4096}{5}} a, \, \, \,  F_8={\frac {3072}{5}}-{\frac {32768}{5}} bt-{\frac {32768}{5}} {t}^{2}+
\frac {262144}{15} {t}^{4},\nonumber\\
F_9&=&-\frac {16384}{15} a, \, \, \,  F_{10}={\frac {8192}{75}}+{\frac {131072}{75}} {t}^{2}, \, \, \, F_{11}=0,\, \, \, F_{12}=\frac {16384}{225}
\end{eqnarray*}

\section*{Acknowledgments}
The authors thank the anonymous referees for their valuable suggestion.
This work is supported by the National Fundamental Research Program
of China (Contact 2011CB921503), the National Science Foundation of
China (Contact Nos. 11274051, 91021021, 11271052).

\end{document}